%% file: grbpaper.tex
\documentclass[a4paper,british,a4paper,fleqn,usenatbib]{mnras}
\usepackage[T1]{fontenc}
\usepackage[latin9]{inputenc}
\usepackage{babel}
\usepackage{array}
\usepackage{varioref}
\usepackage{textcomp}
\usepackage{url}
\usepackage{amsmath}
\usepackage{amssymb}
\usepackage{graphicx}
\usepackage{esint}
\usepackage[authoryear]{natbib}

\makeatletter

\pdfpageheight\paperheight
\pdfpagewidth\paperwidth

\providecommand{\tabularnewline}{\\}

\usepackage{ae,aecompl}
\usepackage{txfonts}
\input{stats.tex}

\input{resultvalues.tex}

\newcommand{\NH}{ {N_{\rm{H}}} }
\newcommand{\NHmajor}{ {N^{\rm{major}}_{\rm{H}} }}
\newcommand{\NHgal}{ {N^{\rm{MW}}_{\rm{H}} }}
\newcommand{\NHMW}{ {N^{\rm{MW}}_{\rm{H}} }}
\makeatother

\begin{document}
\title[GRBs x-ray their host galaxy gas]{Galaxy gas as obscurer:
I. GRBs x-ray galaxies and find a $N_{\text{H}}^{3}\propto M_{\star}$
relation }
\author[Buchner, Schulze \& Bauer]{
Johannes Buchner$^{1,2}$\thanks{E-mail: johannes.buchner.acad@gmx.com}, 
Steve Schulze$^{2,1}$, Franz E. Bauer$^{2,1,3}$
\\
$^{1}$Millenium Institute of Astrophysics, Vicu\~{n}a Mackenna 4860, 7820436 Macul, Santiago, Chile
\\
$^{2}$Pontificia Universidad Católica de Chile, Instituto de Astrofísica, Vicu\~{n}a Mackenna 4860, 7820436 Macul, Santiago, Chile
\\
$^{3}$Space Science Institute, 4750 Walnut Street, Suite 205, Boulder, Colorado 80301
}{

\date{Accepted XXX. Received YYY; in original form ZZZ}
\pubyear{2016}
\label{firstpage}
\pagerange{\pageref{firstpage}--\pageref{lastpage}}

\maketitle
\begin{abstract}An important constraint for galaxy evolution models
is how much gas resides in galaxies, in particular at the peak of
star formation $z=1-3$. We attempt a novel approach by letting long-duration
Gamma Ray Bursts (LGRBs) x-ray their host galaxies and deliver column
densities to us. This requires a good understanding of the obscurer
and biases introduced by incomplete follow-up observations. We analyse
the X-ray afterglow of all \nlong~ \emph{Swift} LGRBs to date for
their column density $\NH$. To derive the population properties we
propagate all uncertainties in a consistent Bayesian methodology.
The $\NH$ distribution covers the $10^{20-23}\text{cm}^{-2}$ range
and shows no evolutionary effect. Higher obscurations, e.g. Compton-thick
columns, could have been detected but are not observed. The $\NH$
distribution is consistent with sources randomly populating a ellipsoidal
gas cloud of major axis $\NHmajor=10^{23}\text{cm}^{-2}$ with $\Rellscat$
dex intrinsic scatter between objects. The unbiased SHOALS survey
of afterglows and hosts allows us to constrain the relation between
\emph{Spitzer}-derived stellar masses and X-ray derived column densities
$\NH$. We find a well-constrained powerlaw relation of $\NH=10^{\RMlinenorm}\text{cm}^{-2}\times\left(M_{\star}/10^{9.5}M_{\odot}\right)^{1/3}$,
with 0.5 dex intrinsic scatter between objects. The Milky Way and
the Magellanic clouds also follow this relation. From the geometry
of the obscurer, its stellar mass dependence and comparison with local
galaxies we conclude that \textbf{LGRBs are primarily obscured by
galaxy-scale gas}. Ray tracing of simulated \emph{Illustris} galaxies
reveals a relation of the same normalisation, but a steeper stellar-mass
dependence and mild redshift evolution. Our new approach provides
valuable insight into the gas residing in high-redshift galaxies.\end{abstract}


\section{Introduction}

Galaxies radiate because of gas condensed into stars and accretion
onto compact objects. Reconciling that emission with the gas actually
present in galaxies is important to understand processes of galaxy
evolution, including the efficiency of star formation and accretion
processes and the in-fall of cosmological gas into galaxy halos. Most
interesting are constraints at the peak of star formation ($z=1-3$,
\citealp{Madau2014}) and at the peak of the accretion history ($z=0.5-3$,
\citealp{Aird2010,Ueda2014,Buchner2015,Aird2015}). At that time,
the gas content in galaxies was probably higher, as indicated by molecular
gas measurements \citep[e.g.][]{Tacconi2013}, the current primary
tracer of galaxy gas at high redshifts.

The gas in galaxies in turn can attenuate the radiation by absorption
along the line of sight (LOS). This allows one to constrain some properties
of the galaxy gas (e.g., reddining in the UV; \citealp{Boquien2013}).
Because only the total LOS absorbing column can be measured, it remains
unclear in individual galaxies at which scales the absorption is acting,
and thus its density remains undetermined. The alternative is to find
a statistical average over many sight-lines which probe galaxies under
random viewing angles. A further challenge is that the underlying
sources must be detected even in galaxies with the largest gas contents,
and the measurement for absorption must remain sensitive at high column
densities.

This work uses Gamma Ray Bursts (GRBs) as beacons inside high-redshift
(primarily $z=0.5-3$) galaxies to estimate the importance of galaxy-scale
gas obscuration. Long-duration GRBs (LGRBs, duration $>2\text{s}$)
are thought to be caused by the death of massive stars \citep{Woosley2006,Hjorth2012},
and thus approximately trace star formation but with a bias towards
metal-poor galaxies \citep[see][for recent reviews]{Kruehler2015,Perley2016}.
LGRB detection via their prompt gamma ray emission avoids detection
biases due to absorption. The afterglow of LGRBs emits X-rays which
can be used to assess the column density through photo-electric absorption,
primarily from electrons in O and Fe atoms along the LOS\footnote{The alternative hypothesis of a He-dominated, ionised absorber \citep{Watson2013}
is discussed extensively in Section~\ref{sub:He-Watson}.}. This work uses the distribution of LOS column densities to reconstruct
the galactic gas of GRB host galaxies at high redshift.

Three major methodological contributions lead up to this work: 1)
The launch of the \emph{Swift} satellite \citep{Gehrels2004} resulted
in a wealth of GRB data with accurate positioning and timely afterglow
spectroscopy. A series of papers by Campana and collaborators already
exploited these data using best-fit column densities \citep{Campana2006,Campana2010,Campana2012}.
2) By 2012 it was appreciated that current redshift determination
practices introduce systematic sample biases that tend to exclude
dusty, massive host galaxies, potentially under-estimating the entire
column density distribution of the GRB population \citep[see e.g.][]{Fynbo2009,Kruehler2011,Perley2013}.
This work thus also investigates the nature of that bias \citep[see also][for a bright sample analysis]{Campana2012}.
3) \citet{Reichart2001} and \citet{Reichart2002} developed a hierarchical
Bayesian inference framework for GRB population analysis which incorporates
the uncertainties of each individual X-ray spectral analysis. We use
this framework to also consistently include GRBs without redshift
information, and provide a more advanced statistical analysis for
potential redshift evolution of the GRB population.

This work begins with describing the statistical framework in Section
\ref{sec:Methodology}. We describe the LGRB samples used in this
work, the data reduction and spectral analysis procedures in Section
\ref{sec:Data-Reduction}. Section \ref{sec:Results-1} presents the
results of the spectral analysis and population properties. There
we begin with empirical models and in turn investigate more physical
models, redshift evolution, host mass dependence and the effect of
redshift-incomplete samples. We discuss our results in Section \ref{sec:Discussion},
putting our results in the context of previous investigation of LGRB
obscurers (\ref{sub:LGRB-obscurer-models}) and comparing our results
to local galaxies. Finally we look inside simulated galaxies to investigate
their gas columns (Section \ref{sub:Obscuration-in-simulated}) before
summarising our conclusions in Section \ref{sec:Summary}.

\begin{figure*}
\begin{centering}
\includegraphics[width=2\columnwidth]{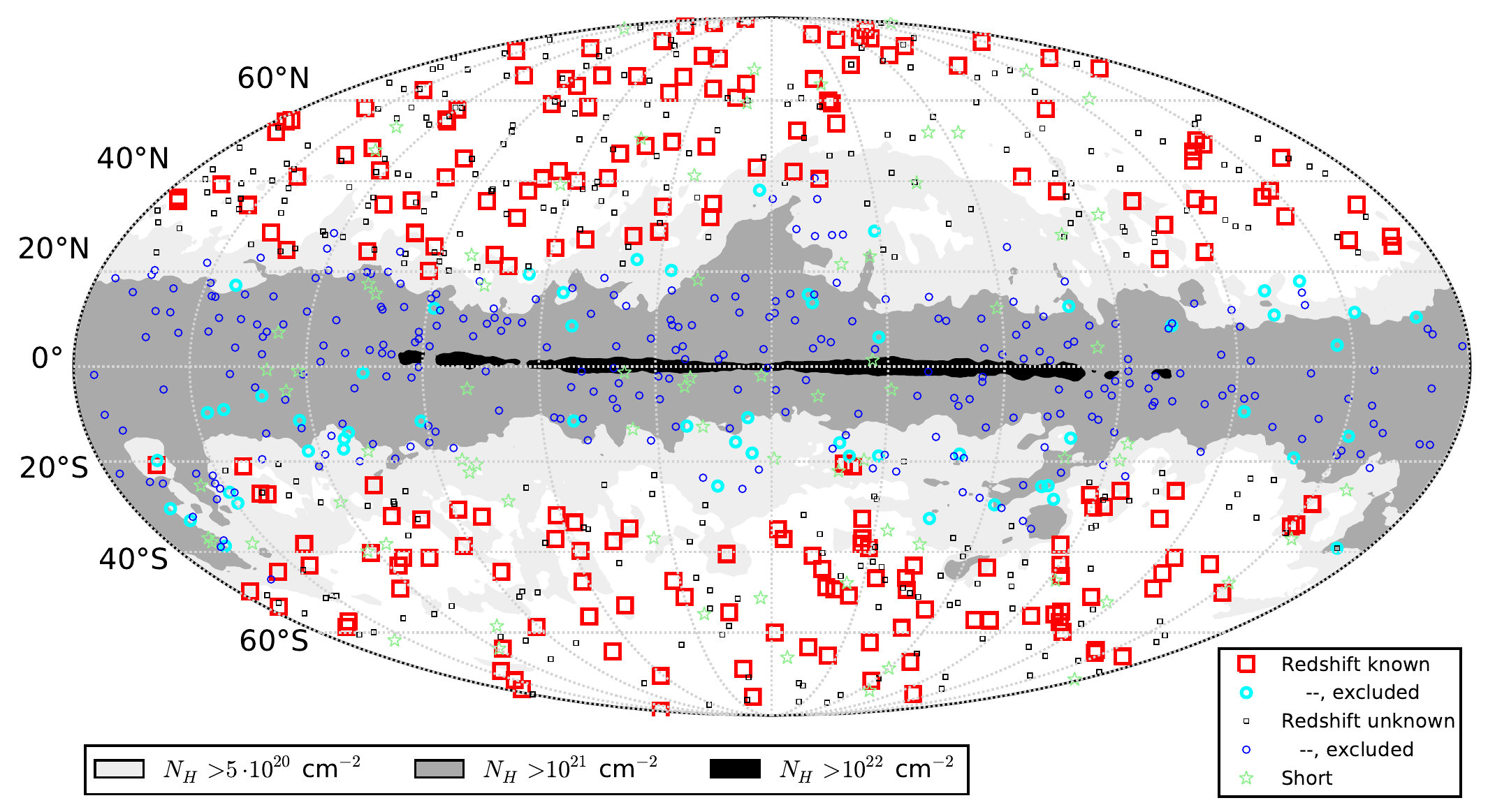}
\par\end{centering}

\caption[Distribution of the GRB sample in Galactic coordinates]{\label{fig:skydist}Distribution of the GRB sample in Galactic coordinates
(points). To avoid Galactic absorption the Galactic latitude $|b|<20\text{\textdegree}$
and regions with $N_{H}>10^{21}\text{cm}^{-2}$ are excluded from
the sample (circles). A smoothened Galactic column density map is
shown in the background \citep{Kalberla2005GalNHdist}.}
\end{figure*}

\section{Methodology}

\label{sub:Population-analysis}\label{sec:Methodology}To analyse
the GRB population we consider several parameterised models. Each
model $M$ predicts the distribution of column densities, $p(\NH|M,\theta)$,
between the GRB and the observer. The parameters $\theta$ of each
model are constrained with X-ray spectral data of a GRB sample, described
in the next section, which is a (Poisson) draw from the general GRB
population. We define the likelihood of observing data $D$ for $n$
objects as 
\begin{equation}
{\cal L}_{M}(\mathbf{\theta})=\prod_{i=1}^{n}\int\,p(D_{i}|\NH,\mathbf{\phi}_{i})\cdot p(\NH,\mathbf{\phi}_{i}|M,\mathbf{\theta})\,d\log\NH d\mathbf{\phi}_{i},\label{eq:like}
\end{equation}
where $p(D_{i}|\cdot)$ is the likelihood of object $i$ to have column
density $\NH$ and other properties $\phi_{i}$ relevant for the X-ray
spectrum, such as redshift and the photon index $\Gamma$ (see below
in Section \ref{sub:X-ray-spectral-analysis}). That likelihood $p(D_{i}|\cdot)$
is discussed in more detail Section \ref{sub:X-ray-spectral-analysis},
but briefly speaking it is a Poisson likelihood due to the count nature
of X-ray photon detections. The sample and its X-ray data are discussed
in the section below.

The distribution of properties within the population is described
by the $p(\NH,\mathbf{\phi}|M,\mathbf{\theta})$, which is the product
of the column density distribution $p(\log\NH|M,\mathbf{\theta})$,
the photon index distribution and further distributions assumed for
the remaining spectral parameters (defined below). The photon index
distribution is assumed to be Gaussian, with the mean and standard
deviation free parameters in $\mathbf{\theta}$ and determined simultaneously.
Models for the column density distribution, which are focus of this
paper, $p(\NH|M,\mathbf{\theta})$ are presented in Section \ref{sec:Results}.

To illustrate the meaning of Equation \ref{eq:like} consider the
case of objects with no information, i.e., a flat $p(D_{i}|\NH)$.
Any population model $p(\log\NH)$ is then equally probable, including
delta functions that predict a constant $\NH$ for all objects. This
follows from $p(\log\NH)$ being normalised in $\log\NH$ space. In
contrast, if an object has a well-constrained likelihood $p(D_{i}|\NH)$,
the integral will be maximised by population models with some probability
there. If two objects have well-constrained, but mutually exclusive
$\NH$ constraints, then the population model $p(\log\NH)$ has to
distribute its probability weight, thereby representing scatter in
the population. Objects without data constraints are effectively ``wildcards'':
their integral mass is concentrated wherever the specific population
model $p(\log\NH)$ is concentrated, creating degenerate solutions
of equal likelihood ${\cal L}$.

Equation \ref{eq:like} defines the likelihood which can be explored
with maximum likelihood methods or a Bayesian approach by varying
the population model parameters $\theta$. This likelihood is well-known
in luminosity function works (e.g., \citealp{Marshall1983,2004AIPC..735..195L,Kelly2008}\footnote{See the Appendix of \citealp{Buchner2015} for a derivation and how
intermediate priors are handled.}) and has also been used in previous studies of the column density
of GRBs \citep{Reichart2001,Reichart2002}. To compare various models,
the Akaike information criterion \citep[AIC;][]{AIC} is adopted.
The AIC prefers the model with the highest maximum likelihood ${\cal L}_{M}(\theta_{\text{max}})$,
but punishes by the number of parameters $m$ according to $\text{AIC}:=2\cdot{\cal L}_{M}(\theta_{\text{max}})-2\cdot m$
(higher AIC is better).

\section{Data}

\label{sec:Data-Reduction}

\subsection{Sample Selection}

\begin{figure}
\begin{centering}
\includegraphics[width=1\columnwidth]{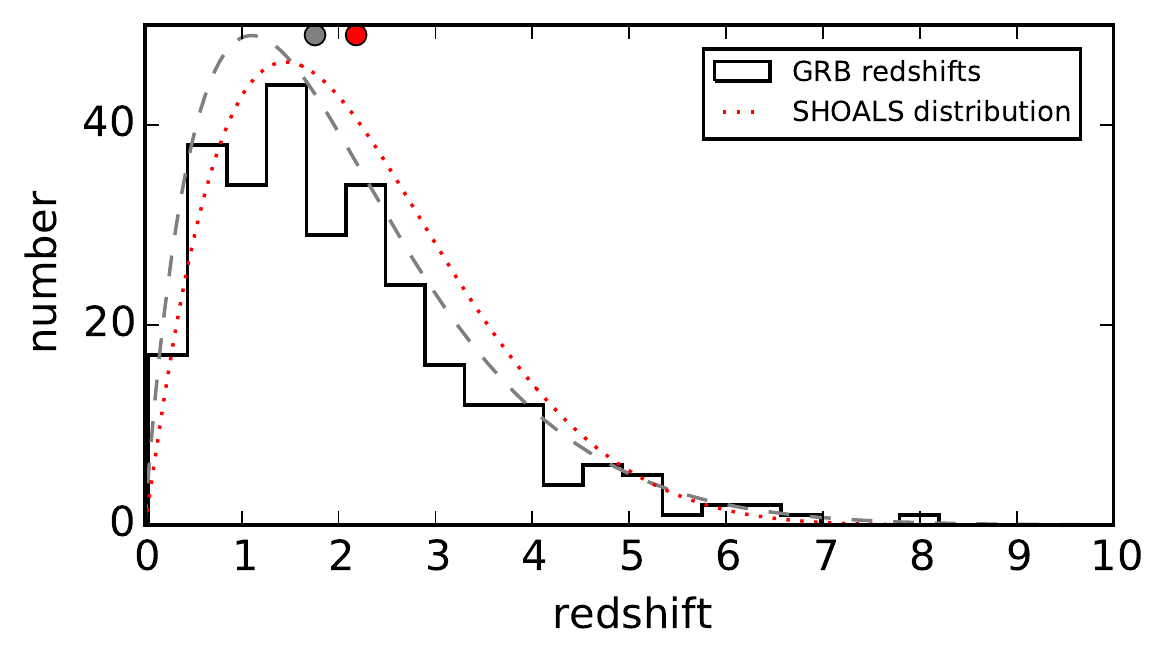}
\par\end{centering}

\caption[Redshift distribution]{\label{fig:zdist}Redshift distribution. The histogram of redshifts
is shown in black. The grey dotted line indicates a fitted Beta distribution.
The red dotted line shows the redshift distribution of the SHOALS
unbiased sample, which peaks at slightly higher redshifts (filled
circles indicate the respective medians). Most LGRBs are found in
the redshift interval $z=0.5-3$.}
\end{figure}

The \emph{Swift} satellite \citep{Gehrels2004} is a dedicated GRB
mission which features on-board detection of GRBs with a wide field
$\gamma$-ray detector and automatic followup using X-ray and optical/UV
telescopes for determining the source position to arcsec accuracy.
We analysed the X-ray spectrum of all GRBs in the \emph{Swift} Burst
Analyser \citep{Evans2010} archive\footnote{\url{http://www.swift.ac.uk/}}
up to June 24th, 2015. Our parent sample is \ntotal~  GRBs detected
by \emph{Swift}.

\begin{table*}
\caption{\label{tab:Sample-selection.}Sample selection.}

\centering{}%
\begin{tabular}{c>{\centering}p{4cm}cc}
\textbf{Sample name} & \textbf{Parent sample} & \textbf{Size} & \textbf{Criteria}\tabularnewline
\hline 
Swift & \url{http://www.swift.ac.uk/}, including non-detected afterglows & \ntotal & \emph{Swift}-detected GRBs\tabularnewline
Long & Swift & \nlong & $T_{90}>2$\tabularnewline
Complete sample & Long & \nlongbrange & $\NHgal<10^{\nhlimit}\text{cm}^{-2}$ and $|b>\blimit\text{\textdegree}$. \tabularnewline
Redshift subsample & Complete sample & \nlongbrangeinzrange & redshift known and $z=0.3-3.2$\tabularnewline
\hline 
SHOALS & - & 119 & \citet{Perley2015a}\tabularnewline
unbiased sample & SHOALS & 119 & $\NHgal<10^{\nhlimit}\text{cm}^{-2}$ and $|b|<\blimit\text{\textdegree}$. \tabularnewline
High-mass subsample & unbiased sample & 25 & \emph{Spitzer} $3.6\text{\ensuremath{\mu}m}$ band $<-22\text{\,mag}$~~~~~~~~~~~~~~~~~~~~~~\tabularnewline
Low-mass subsample & unbiased sample & 94 & \emph{Spitzer} $3.6\text{\ensuremath{\mu}m}$ band $>-22\text{\,mag}$,
or uncertain\tabularnewline
\end{tabular}
\end{table*}

This work ultimately aims to constrain the intrinsic column density
distribution of GRB host galaxies. We select all detected long-duration
GRB, which are associated with the death of massive stars, and therefore
can be expected to trace the galactic gas content through star formation.
To illustrate our sample selection, Figure \ref{fig:skydist} shows
the distribution of all \emph{Swift}-detected GRBs on the sky. The
positions of GRBs are constrained on-board by combining the gamma-ray,
X-ray and optical/UV telescopes. Bursts of short duration (green star
symbols, $T_{90}<2s$) were excluded. These have been identified through
mentions of short duration in associated GRB Coordinates Network (GCN)
Circulars. This leaves \nlong~ LGRBs, indicated as squares and circles
in Figure~\ref{fig:skydist}. Regions where the Milky Way contributes
substantial column densities (over-plotted shades) have to be excluded.
We exclude positions with Galactic column densities $\NHgal>10^{\nhlimit}\text{cm}^{-2}$
and Galactic latitudes $|b|<\blimit\text{\textdegree}$. We call this
sample with \nlongbrange~ objects the \emph{complete sample} (see
Table \ref{tab:Sample-selection.}) as it is unbiased against obscuration.

The availability of redshift information is important to constrain
the obscuring column density from the X-ray spectrum. For \nlongbrangewithz~
LGRBs, redshifts have been determined previously and are indicated
by red squares in Figure~\ref{fig:skydist}. Those \nlongbrangeinzrange~
objects with redshifts in the range $z=0.3-3.2$ are called the \emph{redshift
subsample} (see Table \ref{tab:Sample-selection.}). This criterion
excludes very high-redshift afterglows for which the imprint of absorption
is not observable, and low-luminosity LGRBs at low redshifts which
may have different progenitors or emission mechanisms \citep[see][and references therein]{Dereli2015}.

Whether a redshift has been successfully determined for a particular
LGRB depends on many factors and thus the LGRBs with available redshifts
constitute a biased subsample, showing higher fluxes, lower absorption
than carefully constructed samples with dedicated follow-up \citep{Fynbo2009}.
Working with unbiased samples is thus important to determine the underlying
distribution \citep[e.g.][]{Campana2012}. Such samples are pre-selected
by GRB position relative to the Sun, Moon, galactic plane and available
observatories, but importantly not by afterglow detection or magnitude.
The sample is then followed up with deep ground-based observations
to determine redshifts and host properties \citep[e.g.][]{Jakobsson2006a,Fynbo2009,Greiner2011,Kruehler2012,Jakobsson2012,Schulze2015S,Perley2015a}.
The largest unbiased sample to date is the \emph{Swift} Gamma-Ray
Burst Host Galaxy Legacy Survey \citep[SHOALS,][]{Perley2015a}. Their
redshift distribution is depicted in Figure \ref{fig:zdist} as a
red dotted line. In comparison, the redshift distribution of the complete
sample, where redshifts are available, peaks at lower redshifts than
the redshift subsample (black histogram). 

To overcome the biases of redshift selection, two approaches are followed:
(a) The entire sample is used, including objects without determined
redshift. This sample does then not have a redshift selection bias,
but has low redshift completeness ($\sim40\%$). The distribution
in the sky of this sample is shown in Figure \ref{fig:skydist} with
square symbols. (b) The SHOALS sample is adopted with the same Milky
Way absorption criteria. The resulting 105 objects form an \emph{unbiased
sample} with $\sim90\%$ redshift completeness. For objects with known
spectroscopic redshift we fix the redshift during spectral analysis,
for objects without redshifts or photometric redshifts we adopt as
a redshift prior the unbiased distribution of SHOALS.\nocite{Kruehler2012,Jakobsson2012,Greiner2015}\input{gcncite.tex}

\citet{Perley2015b} investigated the masses of GRB host galaxies.
They obtained \emph{Spitzer} follow-up observations of the SHOALS
sample in the $3.6\mu m$ band, where $-22\,\text{mag}$ corresponds
approximately to a stellar mass of $10^{10}M_{\odot}$. We split our
\emph{unbiased sample} further into \emph{low-mass} and \emph{high-mass
subsamples} using this criterion (see Table \ref{tab:Sample-selection.}).
These subsamples have 94 and 25 objects, respectively.

\subsection{Data reduction}

\label{sub:Data-reduction}X-ray observations were taken using the
XRT instrument on-board \emph{Swift} \citep{Burrows2005} which is
sensitive in the $0.2-10\text{\,keV}$ energy range. To minimise pile-up,
XRT is operated in Window Timing (WT) mode for high-flux GRB afterglows.
Otherwise, Photon Counting mode (PC) is used. GRBs show strong evolution
in their light curve and spectral hardness. This work focuses on the
emission of the X-ray afterglow, which is assumed here to be intrinsically
a powerlaw, and to have a time-invariant spectral shape. However,
the early evolution of an afterglow ($t\lesssim1000\,{\rm s}$) can
be affected by prompt emission. This phase can be easily identified
by its rapid decay ($t^{-\alpha}$ with $\alpha>2$) and spectral
softening. For this we turn to the Swift Burst Analyser, which analyses
the light curve as described in \citet{Evans2009}. Briefly speaking,
the light curve is approximated by piece-wise powerlaw evolutions.
We discard the initial two time intervals if either show a powerlaw
decline with a slope steeper than $-2$, and any immediately following
intervals that also abide by this criterion. Flare intervals are also
discarded. The remaining time segmentation was checked and corrected
by individually inspecting each light curve and hardness ratio. The
final time segmentation is listed for each source in the catalogue
released with this paper (see below).

The spectra from the chosen time segmentation was then extracted using
the Swift Burst Analyser, which automatically screens event files,
selects appropriate energy ranges, applies grade filtering, and chooses
spectral extraction regions to avoid pile-up while maximising S/N.
Background spectra were extracted from appropriate surrounding areas.
Ancillary Response Files (ARFs) and Response Matrix Files (RMFs) were
computed. The $0.5-5\text{keV}$ data were used in the spectral analysis
below, as XRT is most sensitive there with the background well-behaved
and sub-dominant.

Because XRT automatically observes bright sources in WT mode, but
faint sources and late-time observations in PC mode, two data sets
may be available for any GRB. For the majority of GRBs, WT mode is
not or only very briefly used. Brief WT observations were not analysed
here as it is difficult to constrain the background and in practice
they do not improve constraints over the PC mode observations. The
criterion for including WT spectra was that the background spectrum
must contain more than 150 counts. The PC mode data are always used,
if available. 

In some cases, no time interval can be safely used. This can occur
if only the prompt emission is bright enough to be detected. Furthermore
there are sources with no detected afterglow\footnote{Listed separately on the \url{http://www.swift.ac.uk/} website}.
These sources lacking X-ray data were included (if \emph{Swift}-triggered)
in our analysis nevertheless, as they could be heavily obscured, and
comprise part of the unbiased sample.

\subsection{X-ray spectral analysis}

\label{sub:X-ray-spectral-analysis}The intrinsic afterglow spectrum
is thought to be due to synchrotron radiation \citep{Piran2005}.
We model the relevant portion for the $0.5-5\,\text{keV}$ energy
range as a power law $\phi(E)=A\times E^{-\Gamma}$, which is then
photo-electrically absorbed: once by an intrinsic absorption within
the host galaxy $\NH$ (see below), and once with a Galactic absorption
$\NHMW$ using the \texttt{TBABS} ISM absorption model (\citealp{Wilms2000},
with cross-sections from \citealp{Verner1996}). The source model
parameters are thus the normalisation $A$, the photon index $\Gamma$,
and the absorbing column densities $\NH$ and $\NHMW$. The intrinsic
power law is assumed to remain constant with time: hardness ratio
variations where not found (see above) and luminosity variations do
not affect the Poisson fit when we are only interested in the spectral
shape. The background spectrum is empirically modelled as described
in Appendix \ref{sec:bkg-model}. 

Towards Compton-thick densities, here $\NH>10^{24}\text{cm}^{-2}$
for simplicity, effects beyond photo-electric absorption become important.
Such columns in a dense GRB environment or host galaxy gas could block
even the X-ray afterglow emission. Therefore we search for evidence
of high column densities. We adopt the \texttt{SPHERE} model of \citet{Brightman2011a},
which describes photo-electric absorption, Compton-scattering and
line fluorescence computed self-consistently in a spherical, constant-density
obscurer geometry with a powerlaw source in the centre. The \texttt{SPHERE}
model supports column densities up to $\NH=10^{26}\text{cm}^{-2}$.
Solar metallicities \citep{Anders1982} are assumed when deriving
the neutral hydrogen-equivalent column densities $\NH$. However,
LGRBs appear to be often found in low-metallicity environments \citep{Graham2013}.
Derived column densities should thus be primarily considered as metal
column densities as relevant for photo-electric absorption of X-rays.
We also repeated our entire analysis using the spectral analysis the
\texttt{TBABS} ISM absorption model (\citealp{Wilms2000}, with cross-sections
from \citealp{Verner1996}) and find consistent column densities within
the uncertainties. However, in high-obscuration sources ($\NH\gtrsim10^{22.5}{\rm cm}^{-2}$)
\texttt{TBABS} (and other photo-absorption models) sometimes produces
a secondary, Compton-thick solution of lower probability. This solution
is not physical as it is not present when analysing with the more
appropriate \texttt{SPHERE} model. We therefore use the \texttt{SPHERE}
model throughout.

\begin{figure*}
\begin{centering}
\includegraphics[width=1\textwidth]{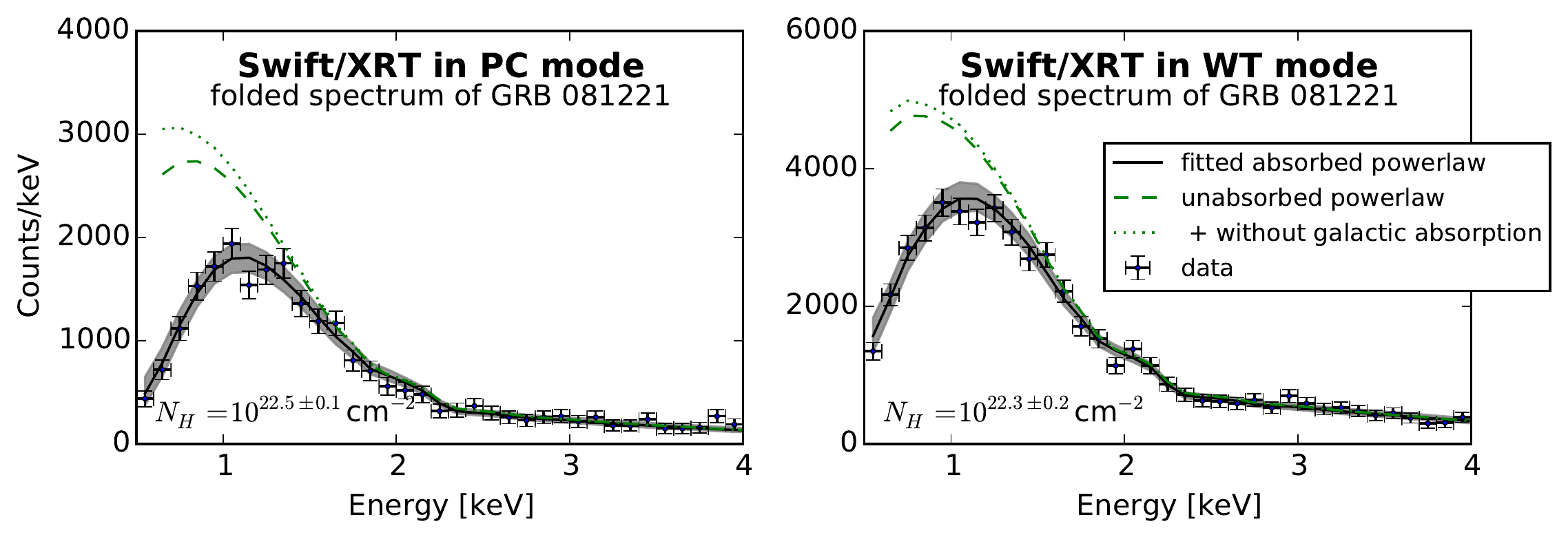}
\par\end{centering}

\caption[XRT X-ray spectrum]{\label{fig:XRT-X-ray-spectrum}XRT convolved X-ray spectrum of GRB~081221.
\emph{Left}: PC mode spectrum. \emph{Right}: WT mode spectrum. The
absorbed powerlaw model (black line, with $3\sigma$ uncertainties
in grey) fits the data (black error bars, binned for plotting) well.
For this source, the column density is $\NH\approx10^{22.5}\text{cm}^{-2}$.
The dashed green line indicates the fit of an unabsorbed powerlaw,
which is clearly ruled out by the low number of counts at $0.5-1\,\text{keV}$.
The dotted line shows the effect if the Galactic absorption, $2\times10^{21}\text{cm}^{-2}$
for this source, is also removed.}
\end{figure*}

To obtain probability distributions for the column density $\NH$,
a Bayesian methodology is adopted for analysing the X-ray spectrum
\citep{Dyk2001,Buchner2014}. The Bayesian X-ray Analysis (\texttt{\textsc{BXA}})
software, which connects the \texttt{\textsc{Sherpa}} X-ray spectral
analysis tool \citep{Freeman2001} to the \texttt{\textsc{MultiNest}}
algorithm \citep{Feroz2009,Feroz2013}, is used with a Poisson likelihood
\citep{Cash1979}. This methodology has the benefit of exploring the
full parameter space and propagating correlated uncertainties, e.g.,
between redshift, obscuring column and the powerlaw slope. 

\begin{table}
\caption{\label{tab:Priors}Parameters of the spectral model and their priors.}

\begin{tabular}{lc>{\raggedright}p{3.7cm}}
\textbf{Parameter} & \textbf{Symbol} & \textbf{Prior}\tabularnewline
\hline 
\hline 
Normalisation & $A$ & log-uniform $10^{-10}-10^{2}\text{keV}^{-1}\text{cm}^{-2}\text{s}^{-1}$\tabularnewline
Powerlaw Slope & $\Gamma$ & uniform $1-3$\tabularnewline
Column density (intrinsic) & $\NH$ & log-uniform $10^{19}-10^{26}\text{cm}^{-2}$\tabularnewline
Galactic column density & $\NHMW$ & log-normal around LAB value; standard deviation $\frac{1}{3}\ln3$\tabularnewline
Redshift & $z$ & fixed if spectroscopy available; otherwise SHOALS distribution (Figure
\ref{fig:zdist})\tabularnewline
\end{tabular}
\end{table}
The Bayesian approach requires the explicit specification of priors.
They are listed for each parameter separately in Table \ref{tab:Priors}.
For the Galactic column density $\NHMW$, a informative normal prior
is adopted around the value measured by the Leiden/Argentine/Bonn
(LAB) Survey of Galactic HI \citep{Kalberla2005GalNHdist} at the
source position. This estimate may be slightly off in unfortunate
conditions, if the Galactic gas is very structured and/or the position
is not precisely known. This uncertainty is allowed by putting a Gaussian
prior around $\log\NHMW$ with a standard deviation of $\sigma=\frac{1}{3}\log3$,
i.e., allowing a three sigma deviation of a factor of 3 in $\NHMW$.
This only broadens our uncertainties in $\NH$, especially when $\NHMW\approx\NH$.
The uninformative priors adopted for $\NH$ and $\Gamma$ are later
replaced by the populations' column density distribution and thus
do not influence the results. For completeness, we include GRBs without
redshift information. For these we adopt as a redshift prior the distribution
of the SHOALS sample (see Figure \ref{fig:zdist}, red dotted line),
which encodes the assumption that these GRBs stem from the same underlying
distribution. \citet{Perley2015a} tested the influence of their fluence
cut on the redshift distribution and found it to be negligible.

Figure~\ref{fig:XRT-X-ray-spectrum} shows an example of a fitted
X-ray spectrum in both PC and WT mode of a highly obscured GRB. None
of the objects show contradictory constraints between the two modes,
which could occur due to poor fits of the source or background spectrum.
As we have analysed the WT and PC mode spectra separately, no assumptions
about the two spectra having the same luminosity or photon index are
made. Finally, the constraints for column density $\NH$ from the
WT and PC mode spectra are merged (if both available) by multiplying
the PC mode probability distributions (in $\NH$ and $\Gamma$) by
the WT $\NH$ probability distribution, thus tightening the constraints
on $\NH$. The photon index $\Gamma$ can show degeneracies with $\NH$
in low-count spectra, so its probability distribution is carried along
as described in Section \ref{sec:Methodology}. The population distributions
of $\Gamma$, $\NH$ and $z$ are constrained simultaneously. The
photon index distribution is reported in Appendix \ref{sec:Gamma-Luminosity}.

\section{Results}

\begin{table*}
\input{cat-excerpt.tex}

\caption{\label{tab:Catalogue}Catalogue (excerpt). Columns: (1) Name. (2)
Duration (-1 indicates unknown). (3) Right ascension (J2000) in degrees.
(4). Declination (J2000) in degrees. (5) Galactic latitude in degrees.
(6) Galactic absorption in $10^{20}\text{cm}^{-2}$. (7) Redshift
(0 indicates unknown). (8) Time selection (in seconds since trigger)
from which the afterglow emission was analysed in X-ray. If ``n/a'',
no interval could be used. If not \emph{Swift} triggered, this is
also noted in this column. (9-12) Derived column density distribution
from the X-ray spectrum analysis, listing (9) the mean in logarithmic
units of $\text{cm}^{-2}$, (10) the 10\% quantile and (11) the 90\%
quantile. This catalogue excerpt contains only the first and last
GRB of our sample, as well as the lowest and highest obscuration GRBs
found.}
\end{table*}
\begin{figure*}
\begin{centering}
\includegraphics[width=2\columnwidth]{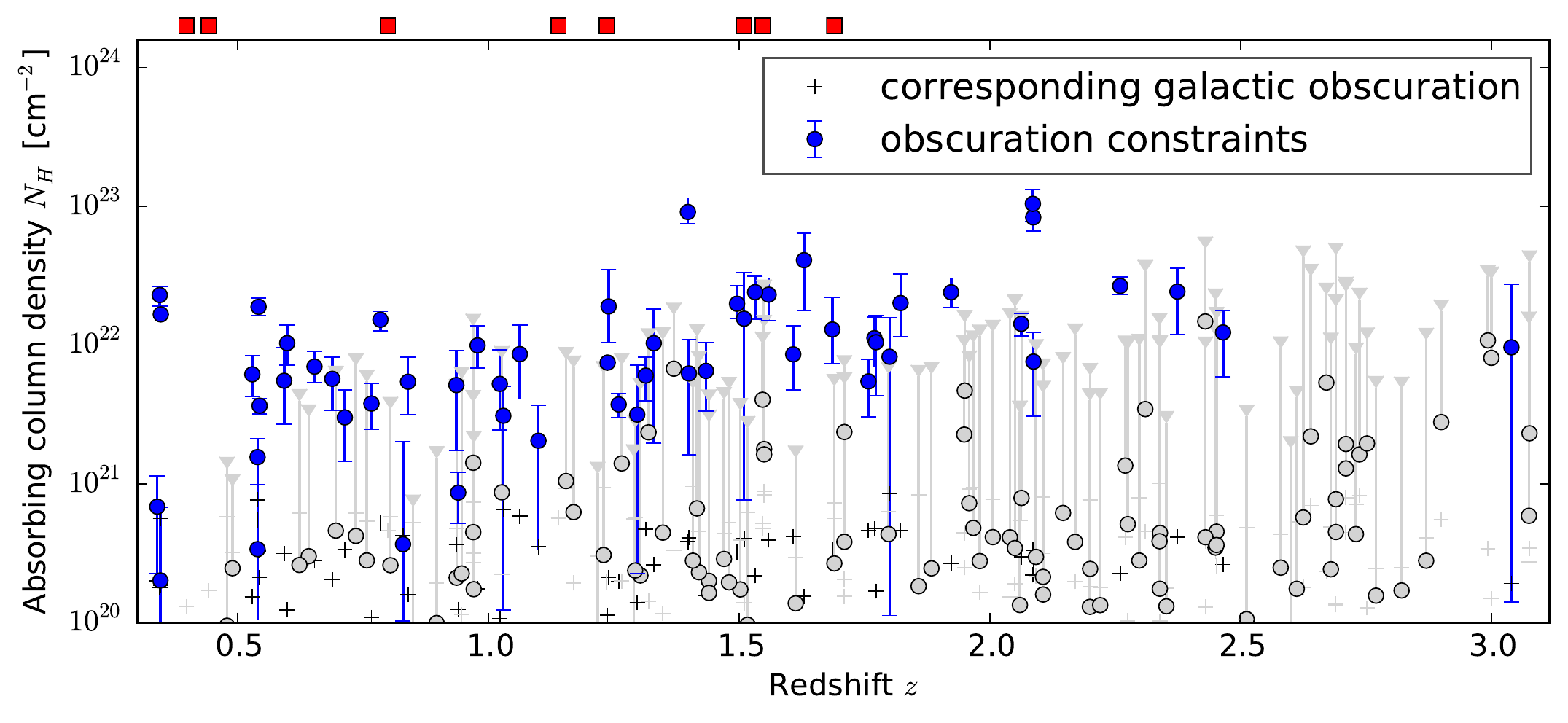}
\par\end{centering}

\caption[Sample distribution in $\NH$ and redshift]{\label{fig:zNH-sample}Redshift $z$ and column density $\NH$ distributions
of the redshift subsample. Blue circles with error bars show objects
with upper and lower limits; the circle indicates the posterior median,
while the error bars show the 90\% posterior probability quantiles.
The black crosses show the Galactic column density corresponding to
each object. Grey circles and crosses show the same, but for unobscured
GRBs which only have upper limits. The red squares at the top indicate
GRBs where no X-ray information is available, either because no afterglow
was detected or no time intervals are free of prompt emission and
flares. They are placed here at an arbitrarily for visualisation,
but in the population analysis all $\NH$ values are considered possible
for these sources.}
\end{figure*}

\label{sec:Results-1}Before inferring the population properties of
GRB obscurers, we briefly describe the spectral analysis results of
the sample. Their distribution in column density and redshift is shown
in Figure \ref{fig:zNH-sample} for the redshift subsample at $z=0.3-3.2$.
Out of the \nlongbrangeinzrange~ objects, \nlongbrangeinzrangesecureobsc~
can be securely identified as intrinsically obscured ($\NH>10^{22}\text{cm}^{-2}$
with $90\%$ probability), and four  can be securely identified as
intrinsically unobscured ($\NH<10^{21}\text{cm}^{-2}$ with $90\%$
probability). GRB~080207 shows the highest obscuration with $\NH\approx10^{23}\text{cm}^{-2}$.
In the complete sample, which comprises all LGRB detections of \emph{Swift},
all sources with X-ray data are constrained to $\NH<2\times10^{23}\text{cm}^{-2}$.
A large portion of the sample only has upper limits for the column
density distribution. The lowest upper limit is $\NH<10^{20.54}\text{cm}^{-2}$
(90\% quantile) in GRB~061021. A catalogue of the sample analysis
results of all GRBs is released with this paper. Its columns are described
in Table \ref{tab:Catalogue}.

To test whether we might have missed any heavily obscured (dark) LGRB,
we simulate PC mode spectra for all sources with $z=1-3$, focusing
on the bulk of the sample redshift distribution. We use the same spectral
parameters (normalisation, photon index) but set the obscuring column
to $\NH=10^{24}\text{cm}^{-2}$. This reduces the median number of
detected counts in the $0.5-5\text{keV}$ range from 10266 (of which
2 are expected to be background counts) to 66 counts. Therefore there
would still be enough contrast to detect and characterise the X-ray
emission of such heavily obscured sources via their extremely hard
spectra, which at these redshifts exposes the FeK$\alpha$ feature
and the low-energy end of the Compton-hump.

\subsection{Empirical Population Models}

\label{sec:Results}\label{sub:Shape-and-evolution}

\begin{table*}
\caption{\label{tab:Empirical-model-params}Empirical models for the column
density distribution for the complete sample.}

\input{paramtable.tex}
\end{table*}
\begin{figure*}
\begin{centering}
\includegraphics[width=2\columnwidth]{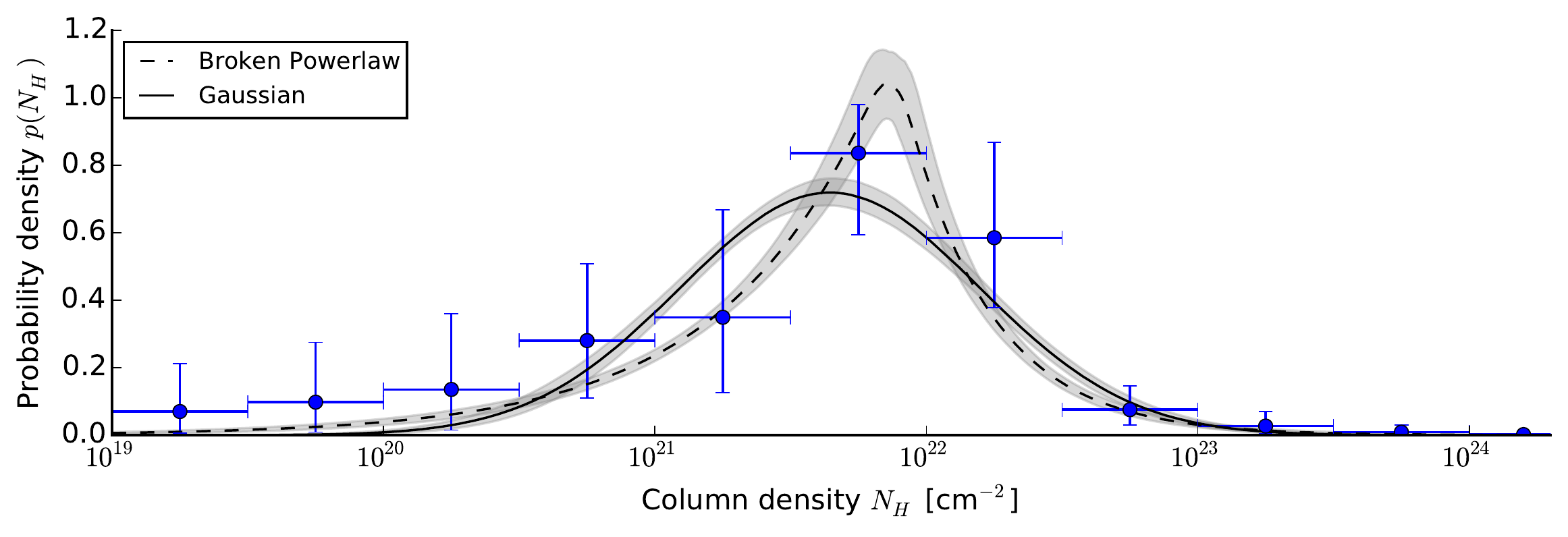}
\par\end{centering}

\caption[Empirical distribution in $\NH$ and redshift]{\label{fig:emp-fits}Column density distributions of \emph{Swift}-detected
LGRBs. Error bars show 2$\sigma$ equivalent quantiles.}
\end{figure*}

To analyse the population properties, specifically the obscurer column
density distribution (CDD), models are adopted which predict the CDD.
To start, we simply want to visualise the data constraints, which
contain large uncertainties and upper limits (shown before in Figure
\ref{fig:zNH-sample}).

Figure \ref{fig:emp-fits} shows the CDD fitted with several models.
The points with error bars are derived by adopting 11 bins (the last
bin extends up to $10^{26}\text{cm}^{-2}$). We find that the column
densities are confined to the $10^{20.5-23}\text{cm}^{-2}$ range.
Two models have been adopted in the literature to describe the CDD
empirically, and are shown in Figure \ref{fig:emp-fits}. \citet{Reichart2002}
and \citet{Campana2010} define a broken powerlaw model (dashed line
fit) as\footnote{In previous works the normalising factor $\ln(10)$ is erroneously
divided, which is important for model comparison.} 
\begin{equation}
p(\NH|M_{\text{BKNPL}},a,b,c)=\ln(10)\cdot\frac{b\cdot c}{c-b}\cdot\begin{cases}
(\NH/a)^{b} & \text{if }\NH\leq a\\
(\NH/a)^{c} & \text{if }\NH>a
\end{cases}.\label{eq:model-bknpl}
\end{equation}
The parameters $b$ and $c$ give the powerlaw slopes at the low and
high-obscuration ends respectively, separated at the break $a$. Alternatively,
a Gaussian distribution of $\log\,\NH$ (solid line fit in Figure
\ref{fig:emp-fits}) has been used \citep[e.g.][]{Campana2010,Campana2012}

\begin{equation}
p(\NH|M_{\text{GAUSS}},\mu,\sigma)=\frac{1}{\sqrt{2\pi}\sigma}\cdot\exp\left[-\frac{\left(\log\NH-\mu\right)^{2}}{2\sigma^{2}}\right].\label{eq:model-gauss}
\end{equation}
The parameters are the centre of the distribution $\mu$ and its width
$\sigma$. 

The constrained parameter values for the two models are listed in
Table~\ref{tab:Empirical-model-params}. The parameters are constrained
by sampling the posterior distribution using \texttt{\textsc{MultiNest}}
\citep{Feroz2009,Feroz2013} through \texttt{\textsc{PyMultiNest}}
\citep{Buchner2014}. Flat priors have been assumed on $\mu$, $\log\sigma$
and $\nu$, $\log a$, $b$, and $c$. In general we find the distribution
to be centred at $N_{H}\approx10^{21.8}\text{cm}^{-2}$ and effectively
spreading two orders of magnitude (see Figure \ref{fig:emp-fits}).
The Gaussian model yields a higher likelihood and since it also has
one parameter fewer, is preferred through a lower AIC value. One possibly
important difference is that the Gaussian has lighter tails (declining
square-exponential) than the broken powerlaw model.

\subsection{SingleEllipsoid: A simplistic physically motivated model\label{sub:SingleEllipsoid:-A-simplistic}}

While we can model the CDD empirically, we ultimately would like to
understand the gas clouds which give rise to the observed column densities.
To this end, we present a simple model of a cloud population, which
could represent the star-forming region the GRB originated in or the
host galaxy. The considered models are gross over-simplifications
of the real scenario, which may include multiple absorbers with sub-structure
and density gradients sampled in a biased fashion by GRBs. However,
as we will show, the simple models are useful to understand the width
and shape of the arising distribution.

\begin{figure}
\begin{centering}
\includegraphics[width=1\columnwidth]{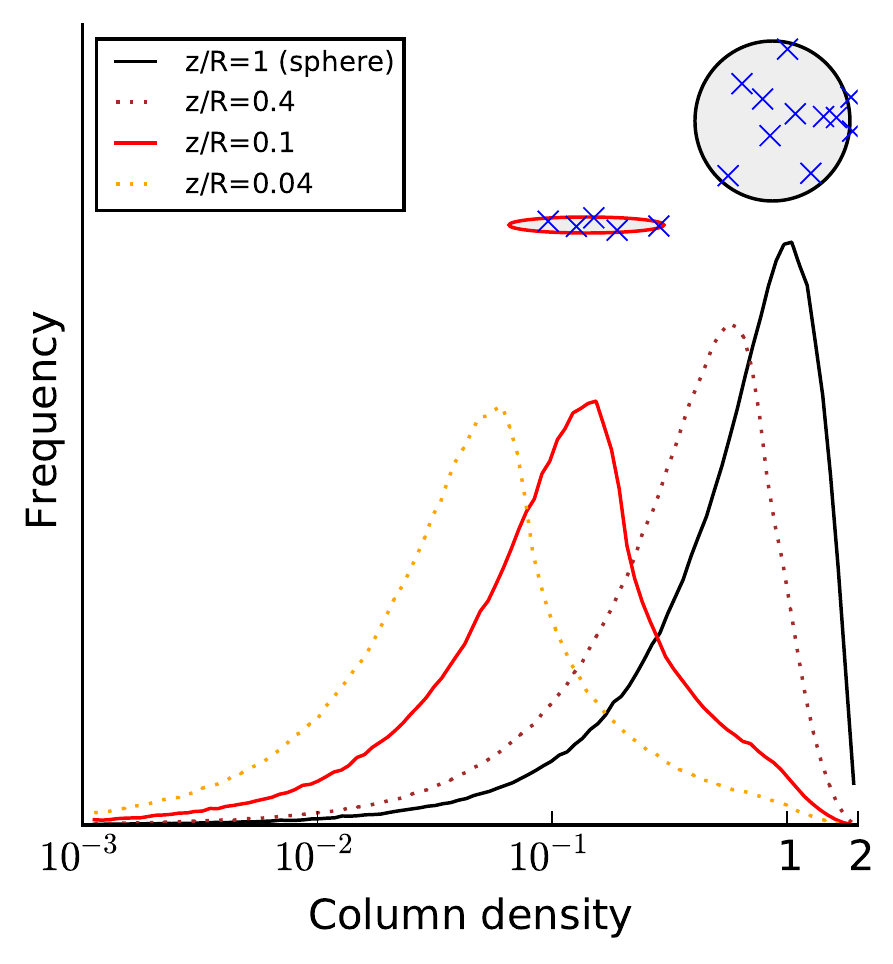}
\par\end{centering}

\caption[Normalised column density distributions of ellipsoids]{\label{fig:ellipsoid}Normalised column density distributions of
ellipsoids. A spherical gas distribution, illustrated in the top right
corner, is sampled with uniformly distributed sources (blue crosses).
The distribution of their column density in random directions is shown
by the thick black line in the plot. In other lines we show the cases
for cylindrical symmetric ellipsoids with flatter height $z$ to radius
$R$ ratios. For instance, the red solid curve shows the case of a
$z/R=0.1$ ellipsoid, also illustrated in red.}
\end{figure}

As an initial toy model, consider a sphere of constant density (radius
1, density 1). For sources distributed uniformly like the gas, the
emerging CDD\footnote{Numerical details of the computation can be found in Appendix \ref{sec:Numerical-Details}.}
is plotted in Figure \ref{fig:ellipsoid} (black solid line). The
largest possible LOS column density is 2 in these units, corresponding
to a full crossing. The most probable column to be observed under
random orientations is around 1, with a long tail down to one order
of magnitude lower. This scenario is illustrated in the top right
corner of Figure \ref{fig:ellipsoid}, with blue crosses indicating
the randomly placed sources. Now consider a flatter geometry, an ellipsoid
of relative height $z/R=0.1$ in cylindrical coordinates, illustrated
in red in Figure \ref{fig:ellipsoid}. The red curve in the plot shows
the corresponding CDD. Here, the distribution is centred at much lower
values, around $0.1$ (i.e., close to the vertical extent), and it
is also very broad, spanning almost three orders of magnitude. 

Such a ellipsoid, representing gas that simultaneously obscures GRBs
and hosts their progenitors, forms the baseline model of our approach
(inspired by, but a generalisation of \citealp{Reichart2002}, see
also \citealp{Vergani2004}). But this model cannot match the data:
the derived CDD in Figure \ref{fig:emp-fits} is broader and less
peaked than those of Figure \ref{fig:ellipsoid}). Actually, it would
be very surprising if it did match, because that would imply that
all gas clouds in which GRBs reside have the same mass and geometry.
We thus define the first physical model to be a population of ellipsoids
with the same height/radius ratio, but with a Gaussian distribution
of total gas densities. The variance of the population in their column
density along the major axis is defined through the parameter $\sigma=\sqrt{\text{var}(\log\NHmajor)}$.
This is mathematically equivalent to convolving the distributions
of Figure~\ref{fig:ellipsoid} with a Gaussian of width $\sigma$.
Subsequently, this model is referred to as the \emph{SingleEllipsoid}
model. The parameters are summarised in Table \ref{tab:Parameters-of-SingleEllipsoid}.
We constrain them from the redshift subsample in the same fashion
as in the previous section.

\begin{table}
\caption{\label{tab:Parameters-of-SingleEllipsoid}Parameters of SingleEllipsoid
model}

\centering{}%
\begin{tabular}{cccc}
Name & Description & Range & Prior\tabularnewline
\hline 
\hline 
$\NHmajor$ & $\NH$ along major axis & $10^{20}-10^{26}\text{cm}^{-2}$ & log-uniform\tabularnewline
$\sigma$ & scatter of $\log\NH^{\text{major}}$ & $10^{-3}-1$ & log-uniform\tabularnewline
$z/R$ & height/radius ratio & $10^{-3}-1$ & log-uniform\tabularnewline
\end{tabular}
\end{table}

\begin{figure}
\begin{centering}
\includegraphics[width=1\columnwidth]{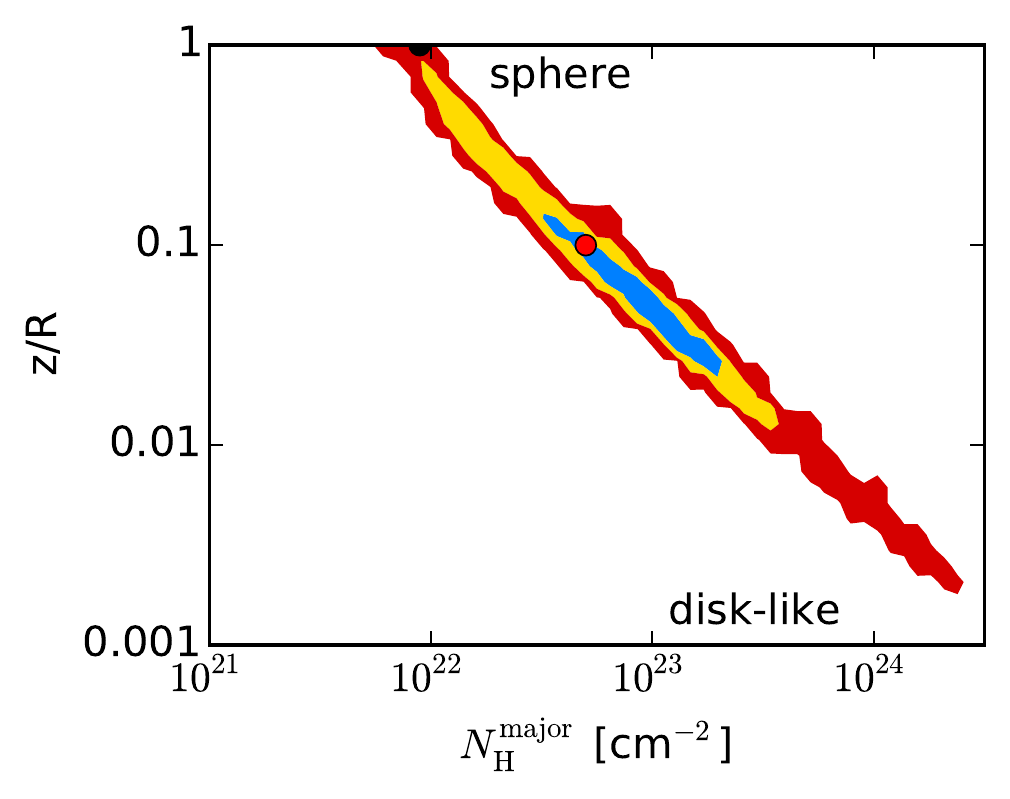}
\par\end{centering}

\caption{\label{fig:Degeneracy}Degeneracy between scale height parameter $z/R$
and the column density along the major axis $\NHmajor$ in the SingleEllipsoid
model. The black and red circle indicate the sphere and disk geometries
illustrated in Figure \ref{fig:ellipsoid}. The contours encapsulate
50\%, 84\% and 99\% of the probability.}
\end{figure}

\begin{figure}
\begin{centering}
\includegraphics[width=1\columnwidth]{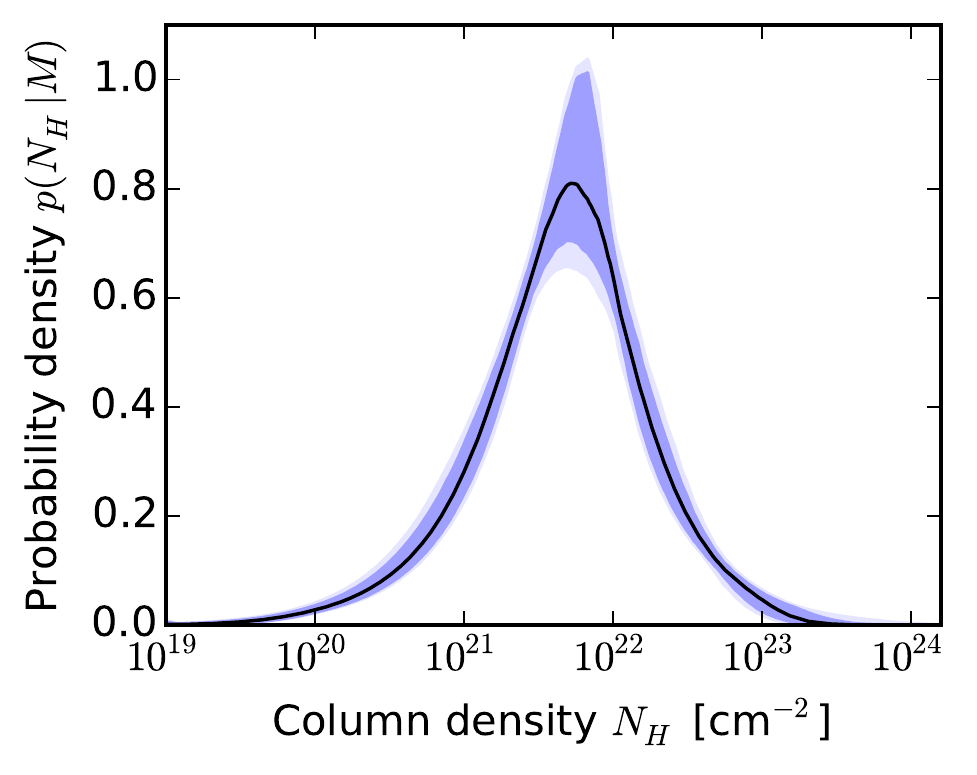}
\par\end{centering}

\caption[Column density distribution of the SingleEllipsoid model]{\label{fig:model-singleellipsoid} Column density distribution of
the SingleEllipsoid model. The dashed line shows the median distribution
from the posterior probability distribution. In dark grey shading,
the $1\sigma$ uncertainty is shown, while light grey shading show
the quantiles encapsulating $90\%$ of the probability distribution.}
\end{figure}

\begin{figure}
\begin{centering}
\includegraphics[width=1\columnwidth]{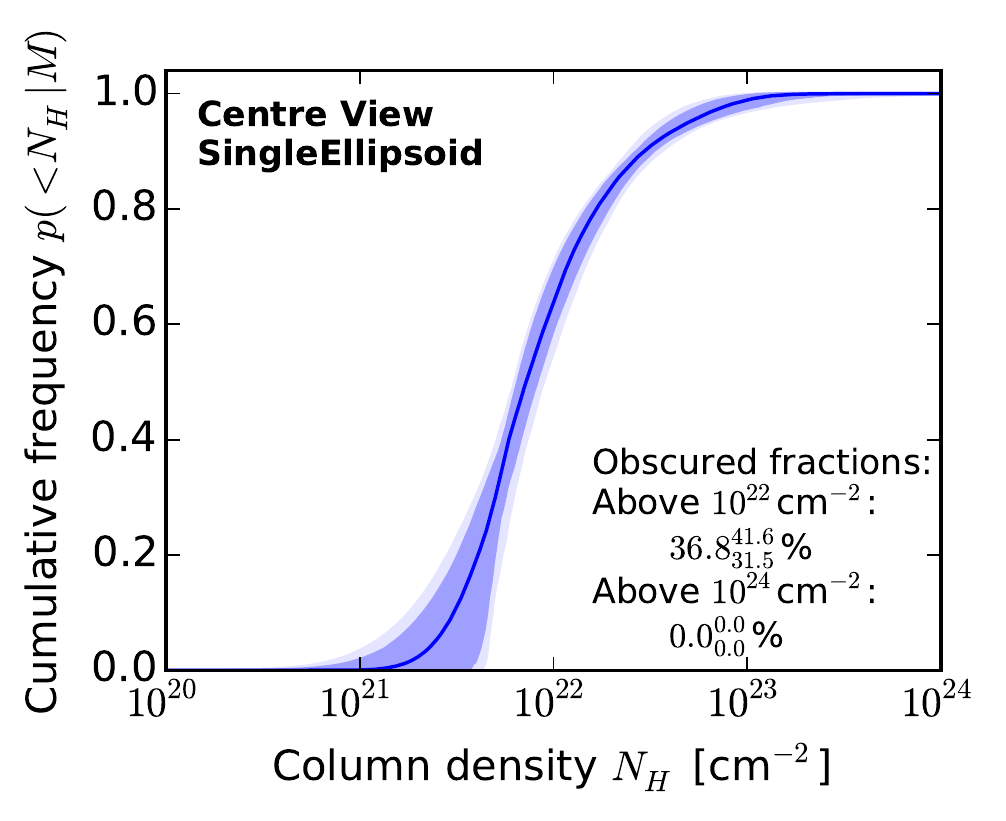}
\par\end{centering}

\caption[Central column density distribution of the SingleEllipsoid model]{\label{fig:model-singleellipsoid-center}Cumulative column density
distribution of the SingleEllipsoid model as seen from the centre.
The solid line shows the median distribution from the posterior probability
distribution. In dark shading, the $1\sigma$ uncertainty is shown,
while light shading represent the quantiles encapsulating $90\%$
of the probability distribution. In this model, about $\sim40\%$
of the sky as seen from the centre is obscured with $\NH>10^{22}\text{cm}^{-2}$,
while there are no Compton-thick lines of sight.}

\end{figure}
The parameters exhibit strong degeneracies between the $z/R$ ratio
and $\NH$ along the major axis, illustrated in the left panel of
Figure \ref{fig:Degeneracy}. The black and red dots indicate the
sphere and disk geometries discussed before in Figure \ref{fig:ellipsoid}.
However, all the possibilities in this degeneracy yield relatively
similar CDDs, shown in Figure \ref{fig:model-singleellipsoid}. The
population scatter $\sigma$ is constrained to $\Rellscat\pm\Rellscaterr$.
This makes the distribution as broad as the empirical model shown
in Figure \ref{fig:emp-fits}. This simple model (SingleEllipsoid)
is already a better fit to the data than the empirical Gaussian mixture
or Broken Powerlaw models, and is also preferred by the AIC. 

\label{sub:Multi-ellipsoid-models}More importantly, however, the
SingleEllipsoid model allows us to derive the CDD as it would be seen
from the centre of the cloud. This is shown in Figure \ref{fig:model-singleellipsoid-center}.
The central CDD spans the $\NH=10^{21}-10^{23}\text{cm}^{-2}$ range,
with no Compton-thick lines of sight. Under this SingleEllipsoid geometry,
up to $50\%$ of the sky would appear obscured with $\NH>10^{22}\text{cm}^{-2}$.

We also investigated possible density gradients by using several co-centred
ellipses, each having free shape and density parameters. Such a model
can, given enough components, reproduce any monotonically declining
density profile. It should also be noted that by embedding a small,
dense component arbitrarily large central column density distribution
can be produced. However, we find that such complications are not
justified by AIC model comparison.

\subsection{Redshift evolution}

We investigate evolutionary trends with redshift by adopting independent
column density distributions in each of five redshift bins,

\begin{equation}
p(\NH,z|M_{X,z},\theta)=\begin{cases}
p(\NH|M_{X},\phi_{1}) & z<0.3\\
p(\NH|M_{X},\phi_{2}) & 0.3<z<1\\
p(\NH|M_{X},\phi_{3}) & 1<z<2\\
p(\NH|M_{X},\phi_{4}) & 2<z<4\\
p(\NH|M_{X},\phi_{5}) & z>4
\end{cases}.\label{eq:model-z-dep}
\end{equation}

In each redshift bin we adopt the Gaussian or Broken Powerlaw empirical
models. This is chosen despite the above finding that the SingleEllipsoid
model is a better fit; the empirical models provide a sufficient characterisation
of the CDD and are faster to evaluate. Their parameters are also easier
to understand and to compare with other works. The constrained parameter
values, using the complete sample, are listed in Table~\vref{tab:Empirical-model-params}.
The last column lists the AIC model comparison values relative to
the single broken powerlaw model (lower is better). The redshift-independent
broken powerlaw is preferred over the redshift-dependent variant.
For the Gaussian model, the redshift-dependent variant is preferred.
This is caused by the significantly lower average column density in
the lowest redshift bin $z<0.3$. Such very local LGRBs are dominated
by low-luminosity afterglows which may be a distinct population \citep{Dereli2015}.
Uncertainties however remain substantial as few LGRBs exist at low
redshifts. At higher redshifts we find no evidence of any redshift
evolution, with the CDD always centred at $\NH\approx10^{21.4-21.8}\,\text{cm}^{-2}$
with small uncertainties. If any redshift evolution exists there,
its effect on the mean column is less than a factor of $\Revolfactor$.
At very high redshifts ($z>4$) the evolution is again uncertain because
the imprint of absorption is redshifted below the X-ray regime. The
uncertainties in Table \ref{tab:Empirical-model-params} indeed highlight
that the strongest constraints come from the $z=0.3-4$ redshift range.
If we replace the model in that range with a SingleEllipsoid model,
the uncertainties of Figure~\ref{fig:Degeneracy} shrink and the
spherical obscurer is ruled out with 99\% probability. We also note
that the redshift-independent models are always preferred, i.e., no
significant redshift evolution is found, when adopting the SHOALS
sample.

\subsection{Host mass dependence of the $\NH$ distribution\label{sub:Host-mass-dependence}}

\begin{figure}
\begin{centering}
\includegraphics[width=1\columnwidth]{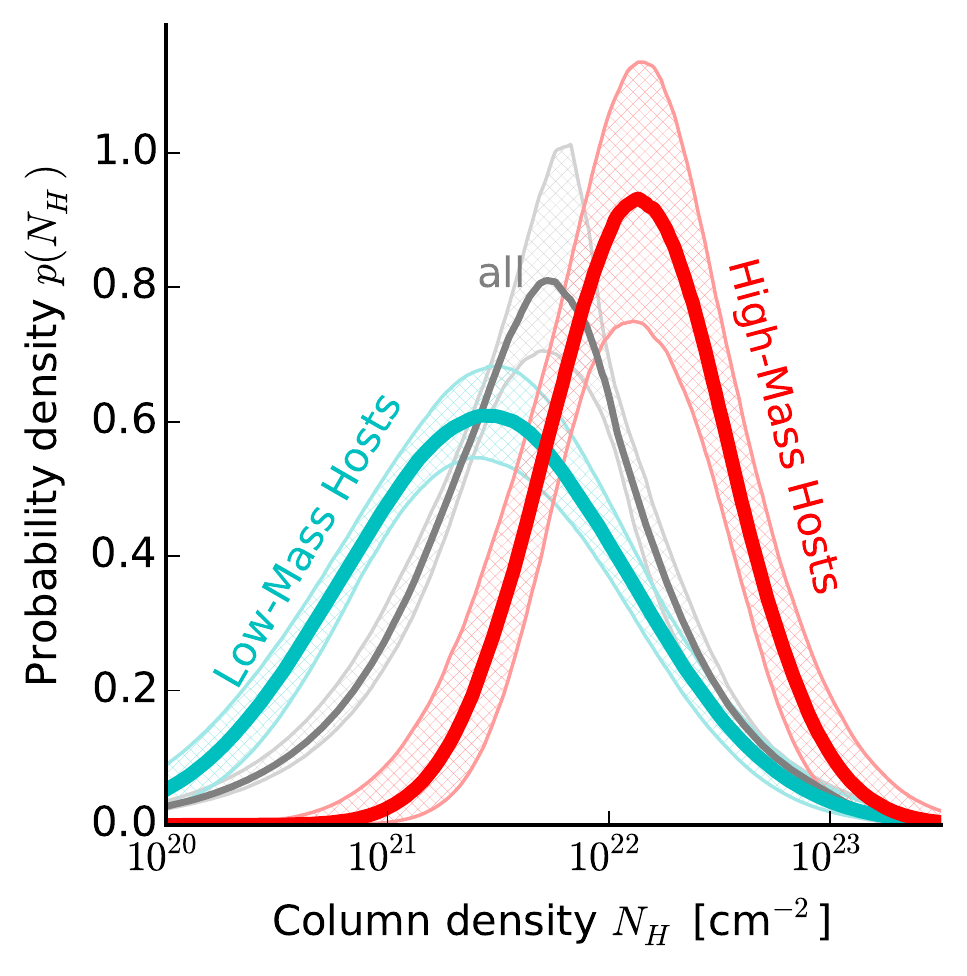}
\par\end{centering}

\caption{\label{fig:Mass-dependence}Mass dependence of the column density
distribution. LGRBs originating in low-mass host galaxies (cyan, $M_{\star}<10^{10}M_{\odot}$)
show lower column densities than those in high-mass galaxies (red,
$M_{\star}>10^{10}M_{\odot}$). For the SHOALS sample \emph{Spitzer}
observations were used to derive stellar masses \citep{Perley2015b}.
In grey, the SingleEllipse model for all LGRBs is shown.}
\end{figure}

\begin{table}
\caption{\label{tab:Gaussian-model-params}Gaussian model parameters for different
samples.}

\centering{}\input{samplestable.tex}
\end{table}
\citet{Perley2015b} investigated the galaxy mass of LGRB hosts and
found that above $M_{\star}=10^{10}M_{\odot}$ virtually all of their
LGRBs had dusty/obscured afterglows. Their SHOALS sample was constructed
only using observability criteria, i.e., quantities that are unrelated
to the host galaxy mass and LGRB obscuration \citep[see][for details]{Perley2015a}.
This subsample selection was then targeted with very deep follow-up
observations to derive redshifts and host galaxy masses. We adopt
their highly redshift-complete sample and split it at $M_{\star}=10^{10}M_{\odot}$
into a low-mass and high-mass subsample (data are described in more
detail in Section \ref{sec:Data-Reduction}). We analysed each subsample
with the Gaussian model. We find that the column densities are drastically
different between the low-mass and high-mass samples, as illustrated
in Figure \ref{fig:Mass-dependence}: The high-mass subsample shows
a five times higher mean column density. In other words, high-mass
host galaxies are preferentially associated with obscured LGRBs, while
a sizeable fraction of low-mass host galaxies have unobscured LGRBs.
The corresponding parameter values are listed in Table~\ref{tab:Gaussian-model-params}.

\subsection{A mass - column density relationship\label{sub:MNH-rel}}

We have now established that the primary driver of the diversity of
LGRB column densities is host galaxy mass, not evolution with redshift.
Consequently we fit a model for the distribution that is stellar mass
dependent. We convert the \emph{Spitzer} $3.4\mu m$ magnitudes given
in \citet{Perley2015b} to masses according to their model. Adopting
instead the relation of \citealp{Meidt2014} does not change our results
significantly.

Figure \ref{fig:MNH-rel} shows our data in stellar mass - $\NH$
space. A clear increase in the column density with mass can be observed,
with no LGRBs in host galaxies of $M_{\star}<10^{9}M_{\odot}$ exhibiting
$\NH>10^{22}\text{cm}^{-2}$, while such sources exist for more massive
hosts. However, there is substantial scatter in the diagram. We first
fit a powerlaw model including a systematic Gaussian scatter for $\log\NH$.
Our fitting method is as before and takes into account the upper limits,
but the CDD model is now mass-dependent. Our relationship can be written
as: 
\begin{equation}
\log\NH=\RMlinenorm+\RMlineslope\cdot\left(\log M_{\star}/M_{\odot}-9.5\right)\label{eq:MNH-rel}
\end{equation}
Figure \ref{fig:MNH-rel} plots the relation (red dashed line) with
its data uncertainties (grey shading). The data uncertainties are
for the intercept $\overline{\NH}=\RMlinenormfine\pm\RMlinenormerrfine$
and the slope $\RMlineslopefine\pm\RMlineslopeerrfine$. The determined
exponent of approximately $\frac{1}{3}$ is noteworthy. The powerlaw
relation of equation \ref{eq:MNH-rel} connects the line integral
$\NH$ on the left with the volume integral $M_{\star}$ on the right
hand side. If both simply scale geometrically with galaxy size ($\NH\sim r$,
$M_{\star}\sim r^{3}$), and the obscuration is primarily due to the
host galaxy, a exponent of $\frac{1}{3}$ is expected. 

We have simultaneously constrained the remaining intrinsic scatter
as a normal distribution. Its standard deviation is $\sigma=\RMlinescatfine\pm\RMlinescaterrfine$
around the relation, shown as a blue error bar on the right of Figure
\ref{fig:MNH-rel}. If instead of a normal distribution we adopt the
SingleEllipse model, the scatter is consistent with zero. In other
words, the observed scatter can be fully explained by the mass distribution
and geometric effects.

\begin{figure}
\begin{centering}
\includegraphics[width=0.99\columnwidth]{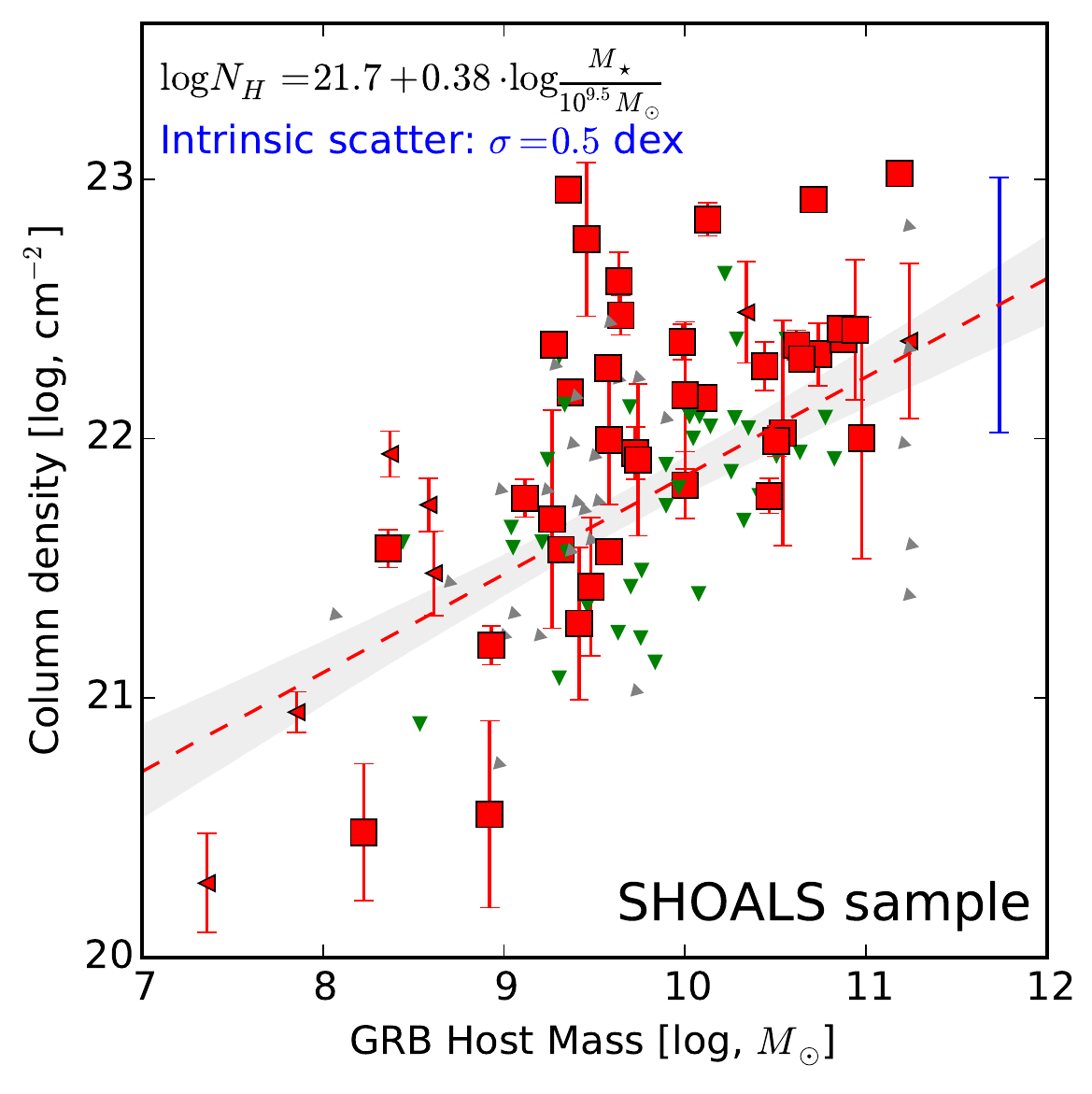}
\par\end{centering}

\caption[Host mass - column density relation]{\label{fig:MNH-rel}Host mass - column density relation. Points are
individual GRBs in the SHOALS survey. Red squares indicate constraints
in mass and column density, green downward pointing triangles are
upper limits in column density, grey triangles indicate upper limits
in both. Columns thicker than $\NH=10^{22}\text{cm}^{-2}$ are only
observed for galaxies more massive than a billion suns. A powerlaw
fit is shown as a red dashed line. Grey indicates the uncertainty
around the slope, while the blue error bar indicates the systematic
scattering around the powerlaw for individual objects.}
\end{figure}

\subsection{Redshift incompleteness bias\label{sub:Redshift-bias}}

Many previous works have only considered LGRBs with determined redshifts,
which is liable to introduce a bias against faint, dust-extincted
hosts. To investigate the nature of this redshift incompleteness bias,
we analyse the redshift subsample, i.e., limit ourselves to LGRBs
with determined redshifts in the $z=0.3-3.2$ range. The derived Gaussian
model parameters are listed in Table~\ref{tab:Gaussian-model-params}.
Compared with the complete sample, the $\NH$ distribution is centred
at lower column densities. This indicates a bias against obscured
LGRBs. In fact, the derived parameter values are most similar to the
low-mass subsample, with the break and low-$\NH$ slope having the
exact same values, and the high-$\NH$ slope falling within $1\sigma$
of the uncertainty. This finding reproduces early works which used
only LGRBs with determined redshifts and found that these only occur
in low-mass host galaxies: The bias on the $\log\NH$ distribution
appears as if only low-mass host galaxies had been selected.

\section{Discussion}

\label{sec:Discussion}

\subsection{The column densities of the GRB population\label{sub:GRB-pop}}

We have analysed the column density distribution of GRBs as probes
of the gas distribution in their host galaxies. We used state-of-the-art
statistical methods to propagate uncertainties in the spectral analysis
and redshift into the population analysis, while remaining careful
of biases from incomplete redshift information.

\begin{figure*}
\begin{centering}
\includegraphics[width=0.85\textwidth]{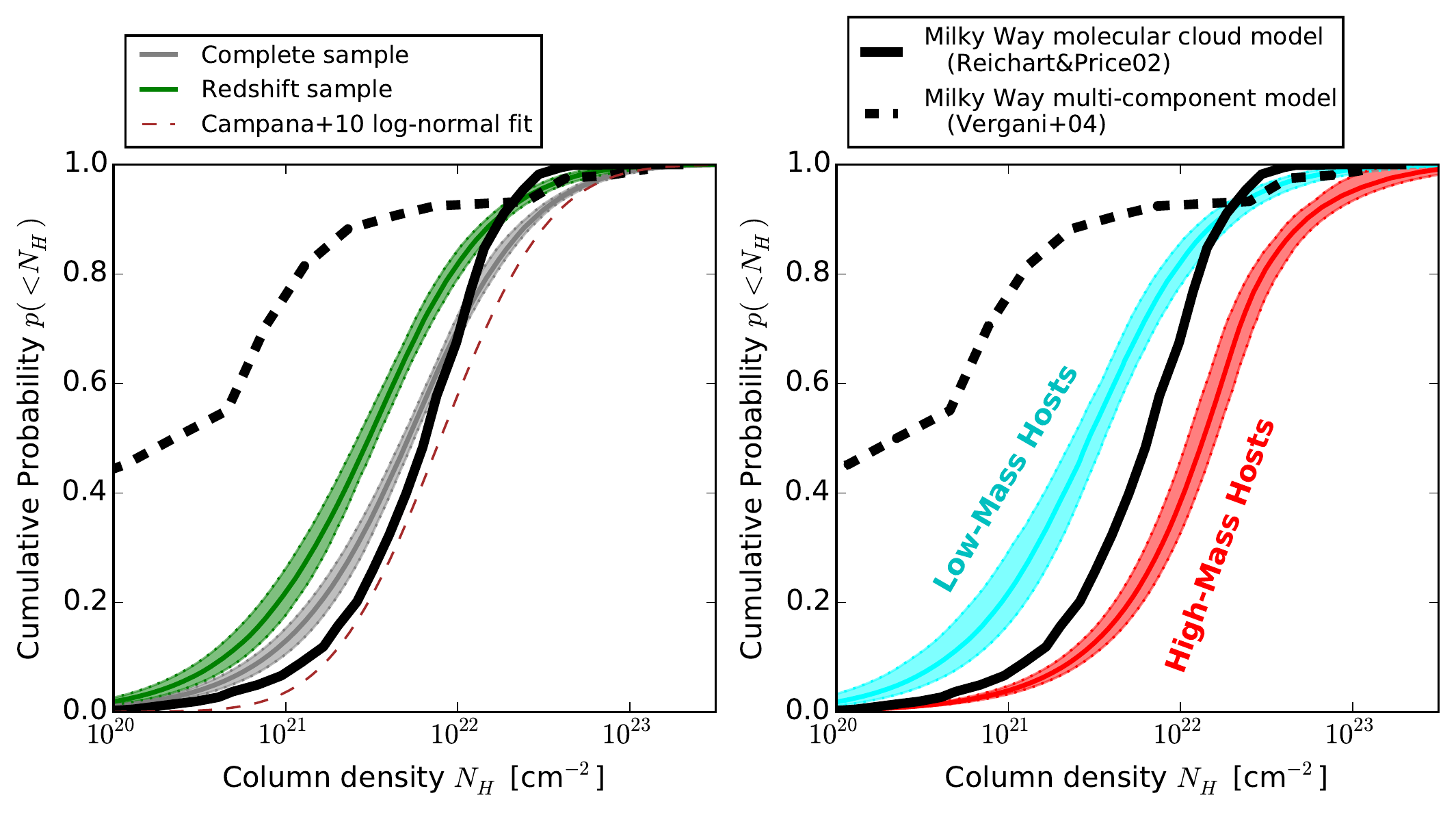}
\par\end{centering}

\caption[Comparison to literature results]{\label{fig:Comparison-to-literature}Comparison to literature results.
Thick solid lines show the models of \citet{Reichart2002} and \citet{Vergani2004}
where GRBs are placed randomly in gas components of the Milky Way;
\citet{Reichart2002} is restricted to molecular clouds. The cumulative
column density distribution from the four samples used in this work
is plotted when adopting the \emph{Gaussian} model (solid: median,
shading: $1\sigma$ quantiles). \emph{Right panel}: Selection of low-mass
(\emph{cyan}) and high-mass (\emph{red}) subsamples show the importance
of the host galaxy. \emph{Left panel}: The complete sample distribution
(\emph{grey}) is very close to the \citet{Vergani2004} model. The
(biased) redshift subsample (\emph{green}) lies systematically at
lower columns by about $0.3\text{dex}$, similar to the low-mass subsample
in the right panel.}
\end{figure*}

In the complete sample, most of the \nlongbrange~~GRBs have column
densities below $10^{21.5}\text{cm}^{-2}$, with the most extreme
spectrum showing $10^{23}\text{cm}^{-2}$ (see Figure \ref{fig:zNH-sample}).
The population can be empirically fitted using a broken powerlaw distribution
which shows a steep decline towards high obscurations (slope of $\sim-1.2$)
and a long tail towards low obscurations (slope of $0.75$), spanning
the $10^{20-23}\text{cm}^{-2}$ range. Thus, heavily obscured (e.g.
Compton-thick) LGRBs, if they exist at all, must be extremely rare.
They have not been seen although XRT is sensitive enough to detect\footnote{In principle, a second class of GRBs could exist behind even more
extreme columns (e.g., $\NH\gtrsim10^{25}{\rm cm}^{-2}$) so as to
render the GRBs undetectable even in the BAT energy band. This would
require a conspicuously bimodal column density distribution which
we do not consider probable a-priori. Such a high column density would
almost certainly have to be local to the GRB.} and characterise them (see Section \ref{sec:Results-1}). Using model
selection we concluded that a better empirical description is provided
by a normal distribution centred at $\log\NH=\Rgaussmean$ with intrinsic
scatter of $\sigma=\Rgaussscat$.

In a series of papers, Campana and collaborators investigated the
column density distribution of LGRBs as a population. \citet{Campana2010}
updated the results of \citet{Campana2006} and analysed a sample
of 93 \emph{Swift}-detected LGRBs with redshift measurements. They
performed a broken powerlaw fit and find the break of the distribution
at $a=21.71_{-0.15}^{+0.14}$, with slopes $b=1.59_{-0.57}^{+1.81}$
and $c=-0.78_{-0.26}^{+0.42}$. This is approximately the same peak
as found in this work, but they find a steeper decline towards low-$\NH$
($b$) and a shallower decline towards high-$\NH$ ($c$) in the population.
This difference is probably due to the handling of the uncertainties
in X-ray spectra and the population analysis. Many sources in their
as well as our analysis show large uncertainties in $\NH$ as derived
from spectral analysis. Errors such as in $2.4_{-1.5}^{+1.7}\times10^{21}\text{cm}^{-2}$
are common, and essentially include the possibility of negligible
intrinsic obscuration. Notably the lower error estimate often includes
values one or two orders of magnitude lower, while the upper error
only doubles the value. In log-space, the best-fit $\NH$ estimator
is thus biased towards the upper limit. This work adopts a Bayesian
methodology to propagate the uncertainties into the population analysis,
which also allows us to treat upper limits consistently.

The obscuration of LGRBs depends strongly on their host galaxies.
LGRBs in high-mass galaxies show higher absorbing columns by $\sim0.7\,\text{dex}$
versus those originating in low-mass galaxies, as shown in Figure
\ref{fig:Mass-dependence}. This agrees with the findings at optical
wavelengths of \citet{Perley2015b}, where massive host galaxies are
virtually always associated with absorbed/dusty afterglows. This suggests
that the obscurer may be primarily the host galaxy itself, with high-mass
galaxies being capable of attracting and holding larger quantities
of gas (see more discussion in the next section). Importantly, this
biases the results when incomplete samples are used: \citet{Campana2012}
noted the bias of \citet{Campana2010}, which appears when considering
only LGRBs where the redshift is determined, as dust-extincted afterglows
are fainter and often are harder to obtain spectra of. They use a
unbiased sample of 58 bright LGRBs and find similar results to \citet{Campana2010},
when comparing a Gaussian fit. This work adopted the SHOALS sample,
which is similar in spirit but larger in size \citep[112 objects,][]{Perley2015a}.
Using newer spectral models which incorporate effects relevant at
high obscuring columns and improved spectral analysis methodology
we are able to make stronger inferences in the derived column density
distribution (Table \ref{tab:Empirical-model-params}). We find that
the bias of considering only LGRBs with determined redshifts is severe,
and that it approximates the exclusion of all massive host galaxies
(see Table \ref{tab:Gaussian-model-params}). The aforementioned effects
can be seen in the left panel of Figure \ref{fig:Comparison-to-literature},
where we compare the redshift sample to the complete sample.

\citet{Campana2010} and \citet{Campana2012} also investigated a
possible redshift evolution of the obscuration. This is interesting
because star forming regions at high redshift, particularly at the
peak of star formation at $z=1-3$, may be more compact. They claimed
that high-redshift GRBs ($z>4$) are more obscured than low-redshift
GRBs. This is based on a KS-test which yields a p-value of 0.08 that
the best-fit column densities are drawn from the same distribution.
P-values are uniform random variates, such that the frequency of yielding
such a result or a more extreme one is high (10\%, but increasing
with the number of tests performed), indicating a substantial probability
of a false positive. \citet{Campana2012} makes more cautious claims
due to the smaller sample size of their unbiased sample. Even if significant,
the best-fit $\NH$ values cannot be drawn from the same distribution
in principle, because the spectral window probed is different. Furthermore,
splitting the sample is problematic because a high percentage of GRBs
have uncertain redshifts. To overcome the limitations of the KS test,
in this work we simultaneously fitted independent distributions in
5 redshift intervals ($z<0.3$, $z=0.3-1$, $1-2$, $2-4$, $z>4$)
and compared their parameters. We find consistent parameters (see
Table \ref{tab:Empirical-model-params}) in the relevant redshift
bins, indicating no redshift evolution around the peak of star formation.
If any redshift evolution of the obscurer is present, it is limited
to modifying the obscuring columns by a factor of $\Revolfactor$,
and thus less important than the host galaxy mass. An exception is
the $z<0.3$ redshift bin, which shows lower obscuration on average.
This may be explained by a dominant low-luminosity GRB population
in that redshift range, which form a distinct population \citep{Dereli2015}.
Alternatively, it could be a side-effect of galaxy-mass downsizing,
which is more pronounced in GRBs \citep{Schulze2015S,Perley2015b}
than in the general galaxy population \citep[e.g.][]{Fontanot2009}.

\subsection{LGRB obscurer models}

\label{sub:LGRB-obscurer-models}

\citet[RP02 hereafter]{Reichart2002} developed a obscurer model based
on the distribution of molecular clouds in the Milky Way. Their mean
radial column densities are $\NH^{\text{major}}\approx10^{22}\text{cm}^{-2}$
with a scatter of $0.2\,\text{dex}$ in their population. Such a cloud
distribution, when including random placement and orientation in such
clouds, was found to be consistent with observations in the analysis
of RP02 with 15 GRBs, and also in the analysis of \citet{Campana2006}
and \citet{Campana2010} which included \emph{Swift} observations.
\citet{Vergani2004} developed a multi-component gas model of the
Milky Way and simulated the LGRB column density distribution with
ray-tracing. They however assumed that a large portion of LGRBs may
occur in diffuse gas. This includes the disk, leading to a high percentage
of LGRBs with low column densities. \citet{Campana2006} ruled out
that model based on their derived column density distribution, and
concluded that LGRBs likely originate in molecular clouds (the remaining
model). A limitation of the RP02 molecular cloud model is that it
is based on the Milky Way, which is atypical in mass and metallicity
for LGRB host galaxies. In this work we developed a more general approach
by deriving the properties of the LGRB obscurer population from the
data.

We find that the column density distribution of LGRBs can be well-described
by a simple model: a single gas component of uniform density, in which
LGRBs are randomly located. The geometry and major axis column density
of the cloud population were tentatively constrained (see Figure \ref{fig:Degeneracy})
to a flat disk with a height-to-radius ratio of $1:20$ and a major
axis column density of $\NHmajor\approx10^{23}\text{cm}^{-2}$. To
explain the broadness of the column density distribution, the best-fit
model has a scatter in $\log\NHmajor$ of $\sigma\approx\Rellscat\pm\Rellscaterr$
(the 10\% quantile is at $\Rellscatqlo$). For comparison, the RP02
model used an essentially flat distribution of scatter $\sigma=0.2\,\text{dex}$
between molecular clouds in the Milky Way. The clouds in the RP02
model have a density gradient, are spherical and therefore match the
data with a lower maximal column density. The right panel of Figure
\ref{fig:Comparison-to-literature} shows the RP02 Milky Way model
in comparison to our results using the high-mass and low-mass subsamples,
which probe stellar masses comparable to, and below that of the Milky
Way. The RP02 model falls in between the constraints from those samples,
implying that LGRBs in Milky Way-size galaxies are more obscured than
what the RP02 Milky Way model predicts. Figure \ref{fig:Comparison-to-literature}
also compares the multi-component model of the Milky Way gas components
by \citet{Vergani2004}. That model assumes that GRBs can also originate
in the atomic hydrogen of the thin disk, which leads to many GRBs
with low column densities. This model is clearly ruled out, leaving
the origin of GRBs in galactic molecular clouds as a plausible scenario.
In that case however the column density of the molecular clouds would
have to increase with galaxy host mass, which is not the case in nearby
galaxies \citep{Larson1981,Bolatto2008,Lombardi2010}. The geometry
constraints suggest another possibility however: giant molecular clouds
could be arranged in a relatively flat disk in which the GRBs are
produced. The number of clouds (and thus the major axis density) should
then scale with the size of the galaxy, as more massive galaxies can
hold larger quantities of gas. This scenario of the galaxy acting
as the primary obscurer appears more likely due to the $\NH^{3}\propto M_{\star}$
relation found (Equation \ref{eq:MNH-rel} in Section \ref{sub:MNH-rel}).

\subsection{Local Galaxies as Obscurers}

\begin{figure}
\begin{centering}
\includegraphics[width=1\columnwidth]{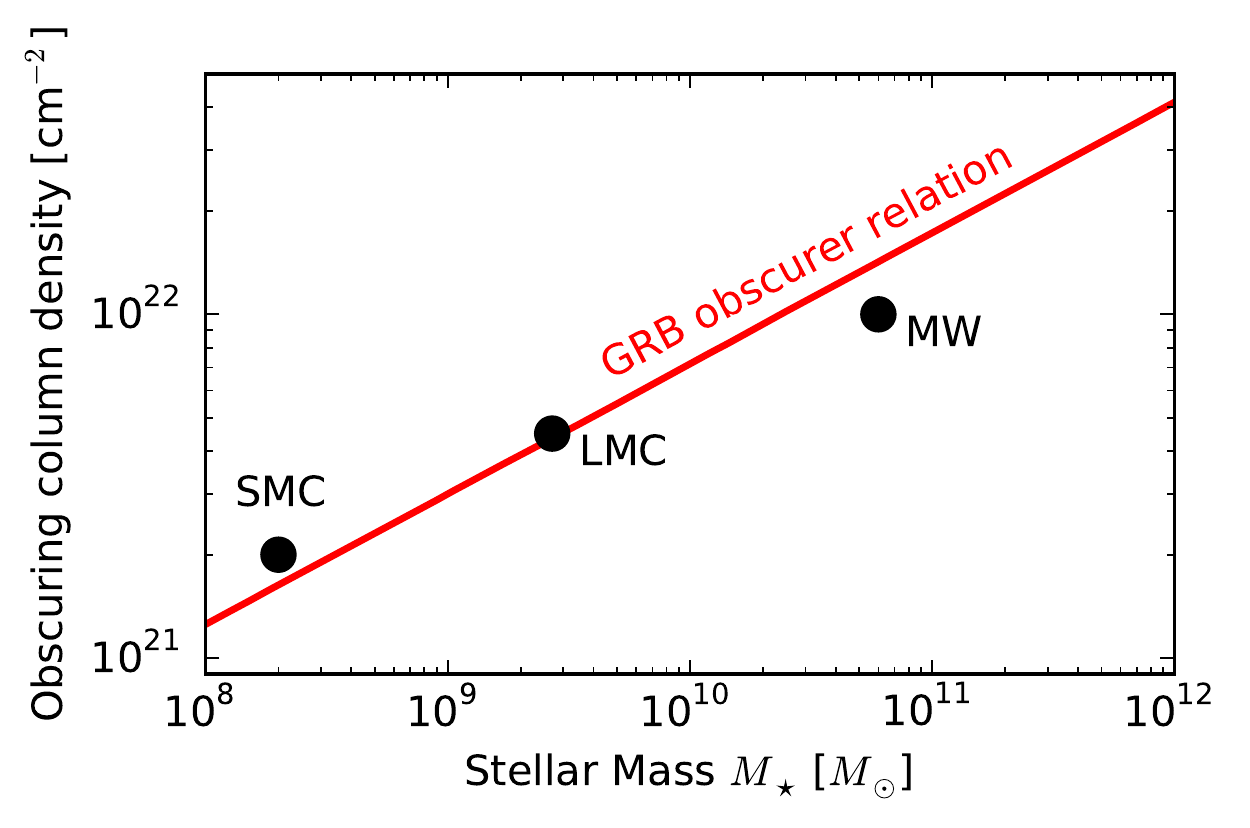}
\par\end{centering}

\caption{\label{fig:localgals}Column densities of local galaxies. We derived
the densest column densities seen in the LMC and SMC, by taking the
99\% highest values from HI radio maps and correcting for metal abundances
to derive a $\NH$ as would be seen by X-ray observations. For the
Milky Way (MW) we use the map of \citet{Dickey1990}. The obscuration
of galaxy gas of each of these local galaxies falls exactly on the
relationship we derive from GRB obscuring column densities (black
line).}
\end{figure}

We verify the mass dependence of the obscurer by determining how well
local galaxies act as obscurers. The Milky Way, the Large Magellanic
Cloud (LMC) and the Small Magellanic Cloud (SMC) cover a large stellar
mass range and their column density of $\NH$ is well mapped. Our
relationship depicted in Figure \ref{fig:MNH-rel} predicts that at
Milky Way stellar masses, $M_{\star}=6\times10^{10}M_{\odot}$ \citep{McMillan2011},
the average galaxy should have some LOS column densities above $\NH>10^{22}\text{cm}^{-2}$.
Galactic column density maps of the Milky Way, as depicted in Figure
\ref{fig:skydist}, show that a small fraction of the sky ($\sim1\%$)
is indeed obscured with $\NH>10^{22}\text{cm}^{-2}$ as seen from
our vantage point. The fraction may be larger from more central regions
of the Galaxy, or from the vicinity of blue, star forming regions
where LGRBs are typically found \citep{Bloom2002,Fruchter2006}. The
Milky Way however lies at the massive extreme when compared with the
host galaxies of LGRBs. The LMC is perhaps a more appropriate galaxy
to consider, with a stellar mass of $M_{\star}\sim3\times10^{9}M_{\odot}$.
Its observed LOS column density reaches values up to $\NH\approx3\cdot10^{21}\text{cm}^{-2}$,
with the star-forming region 30~Dor reaching $\NH\approx9\times10^{21}$
\citep{Bruens2005}. Using the ellipticity ($\epsilon\approx0.3$)
and inclination ($i\approx30\text{\textdegree}$) of the LMC \citep[see][for a review of various measurements]{vanderMarel2006}
the major axis column density should be lower than the observed column
density by a factor of $60\%$. Therefore, from the centre of the
LMC the entire sky has LOS column densities below $\NH<10^{22}\text{cm}^{-2}$.
That is again consistent with our $\NH^{3}\propto M_{\star}$ relationship,
which predicts that such sight lines should be very rare for LMC-mass
galaxies. Finally we consider the SMC, with a stellar mass $M_{\star}\sim3\times10^{8}M_{\odot}$
\citep{vanderMarel2009}. While that galaxy has a smaller HI mass,
its $N_{\text{HI}}$ is higher than in the LMC, reaching $\NH=10^{22}\text{cm}^{-2}$
\citep{Bruens2005}. One should keep in mind that for the SMC and
LMC we relied on HI columns derived from radio observations, whereas
everywhere else in this paper we consistently use X-ray determined
metal columns converted into $\NH$ assuming solar abundances. Therefore
the $\NH$ values we should compare with for the SMC and LMC -- their
metal columns -- actually should be lower given their low gas metallicity
of 0.2 and 0.5 solar, respectively \citep{Tchernyshyov2015}. Taking
the metal abundance into account, we contrast these three local galaxies
(MW, LMC and SMC) in Figure \ref{fig:localgals} against our relationship
and find excellent agreement.

\subsection{A Helium-dominated, circum-burst absorber?}

\begin{figure}
\begin{centering}
\includegraphics[width=0.99\columnwidth]{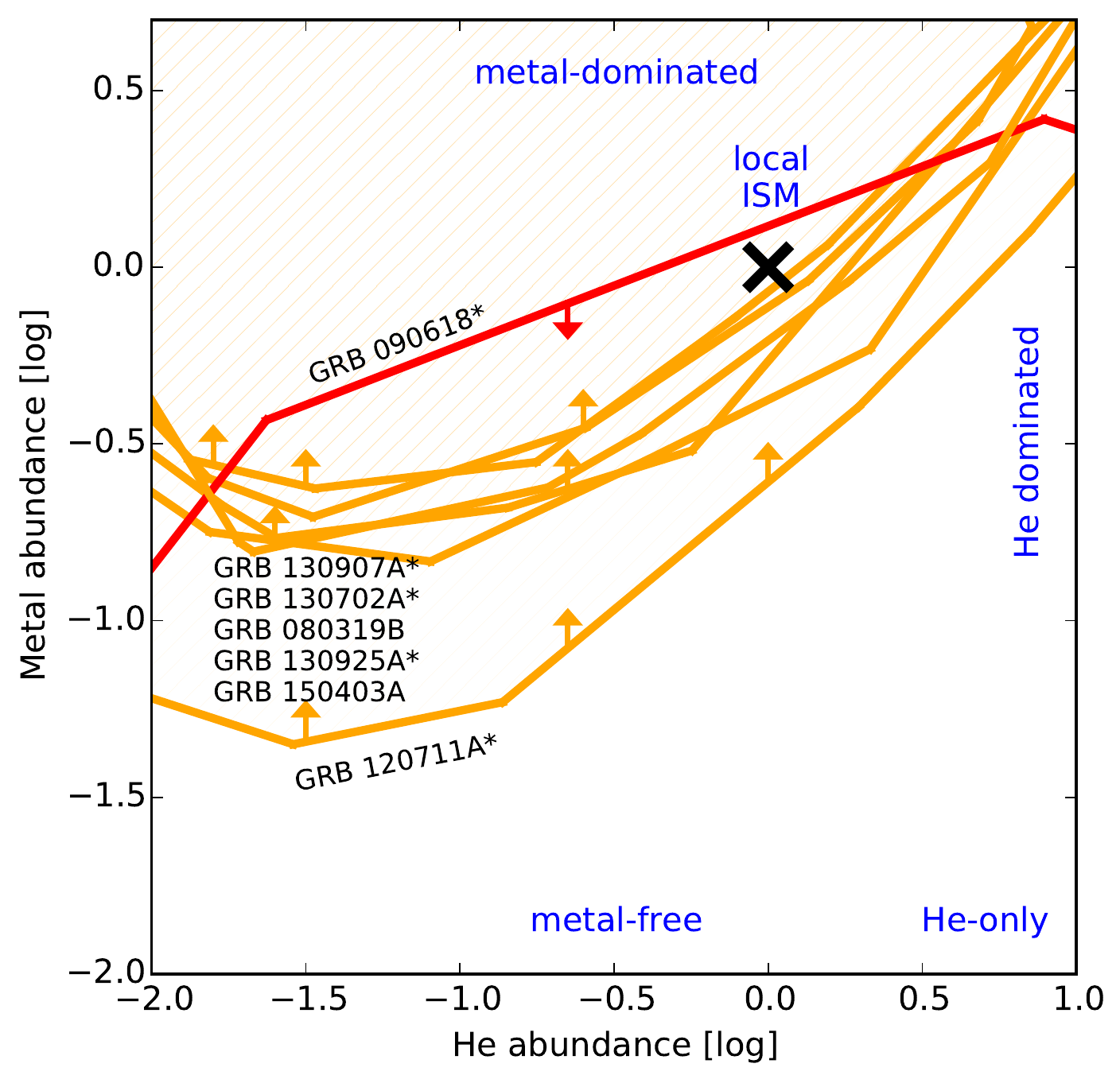}
\par\end{centering}

\caption{\label{fig:abundance}LOS metal abundances for seven LGRBs where the
He and metal abundances could be constrained. The 3-sigma contours
(orange thick lines) of six spectra exclude metal-poor and He-abundant
solutions. For GRB~090618 (red thick line), metal-rich abundances
are excluded. All shown LOS's are consistent with the abundances of
the local ISM \citep[cross,][]{Wilms2000}. A metal-free, Helium-dominated
absorber can be rejected at high significance. LGRBs marked with asterisks
show super-galactic $\NH/A_{{\rm V}}$ ratios.}
\end{figure}
\label{sub:He-Watson}An open issue in the understanding of LGRB afterglow
is the inconsistency between optically and X-ray-derived absorption.
The rest-frame UV extinction, $A_{V}$, and the X-ray derived column
density, $\NH$, show a tendency toward lower-than-galactic ratios
and broad scatter \citep[e.g.][]{Schady2011,Watson2013}. This can
be caused by deviations in abundances, dust-to-gas ratios and/or ionisation
states compared to the galactic ISM. To resolve this inconsistency
goes beyond the scope of this work. However, we study one recently
proposed explanation in detail. \citet[W13 hereafter]{Watson2013}
postulated that gas in the vicinity ($<30\text{pc}$) of the burst
provides the bulk of the X-ray absorption. The powerful burst emission
can destroy dust, fully ionise all hydrogen atoms (which have a low
ionisation energy) and O and Fe atoms (which are few in number) along
the LOS. These atoms would therefore not absorb X-rays. However, for
certain luminosities, not all He atoms would become ionised, because
of their large number. W13 showed that a He-dominated X-ray absorption
spectrum is observationally indistinguishable from one with local
ISM abundances in typical X-ray spectra.\nocite{Sakamoto2008,Romano2006,Sakamoto2009,Rossi2011,vonKienlin2014,Gruber2014,2013GCN..15260...1G,2010GCN..11015...1V,2011GCN..12663...1G,2012GCN..13736...1G}\input{gcncite2.tex}

\begin{figure}
\begin{centering}
\includegraphics[width=0.99\columnwidth]{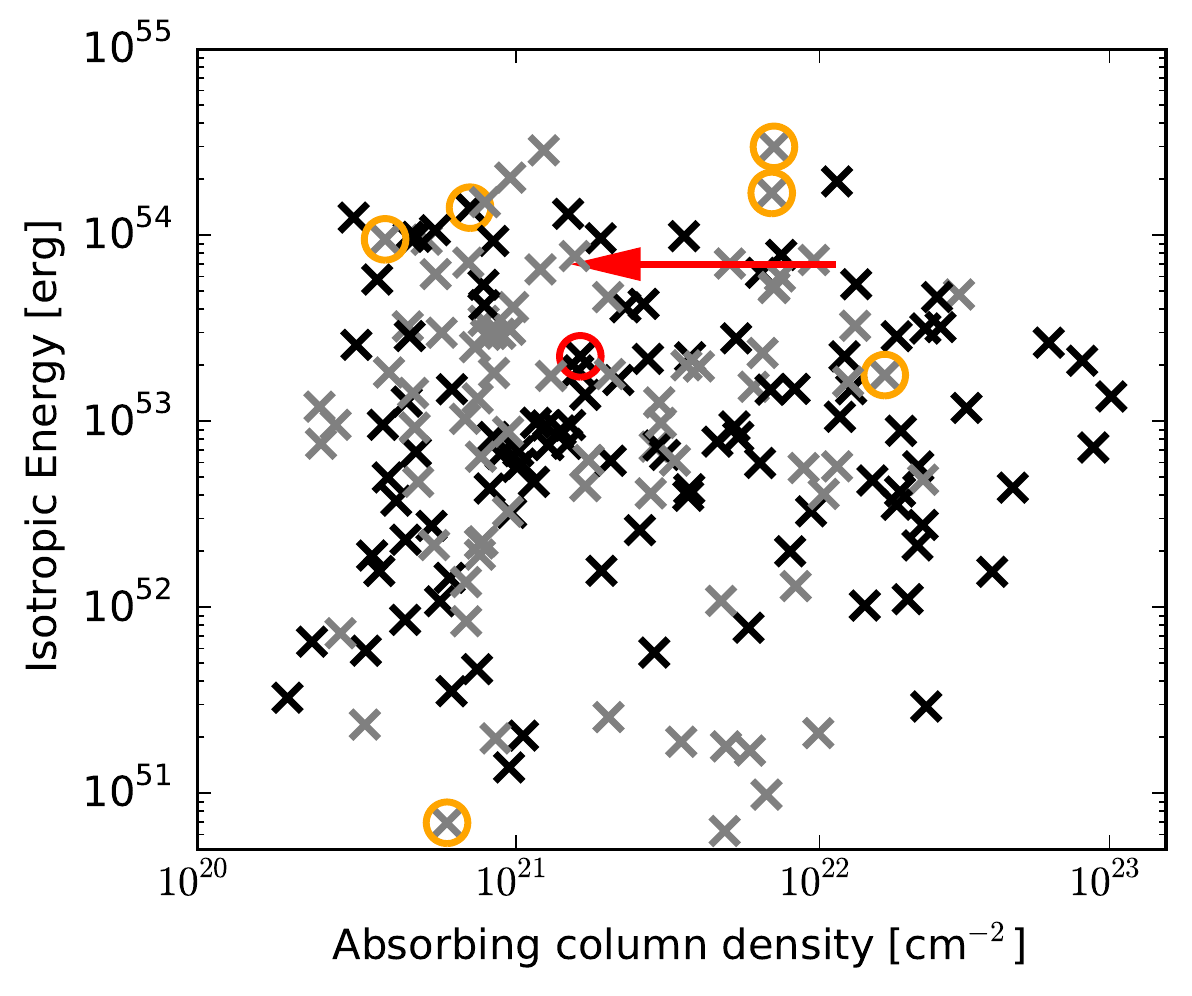}
\par\end{centering}

\caption{\label{fig:energy-NH}The relation between obscurer and LGRB energy.
We plot the effective column density (neutral absorber with local
ISM abundances) against the isotropic energy for sources where the
redshift is known. Black crosses are from the SHOALS sample, grey
crosses denote other LGRBs where the X-ray afterglow spectrum exceeds
1000 counts. The circles indicate the same objects shown in Figure~\ref{fig:abundance},
where there is evidence of metals in the X-ray absorption. The red
arrow indicates a reduction in the effective column density by a factor
of seven if dense circum-burst material were ionised by highly energetic
bursts. However, no such systematic shift is apparent when comparing
to lower energy bursts.}
\end{figure}
In a few cases where photon statistics are robust, the effective abundance
of metals and helium can be constrained directly. We set the helium
and metal abundances each as free parameters in our fit (\texttt{TBVARABS}
model, \citealp{Wilms2000}), and analysed all LGRB XRT spectra with
more than 5000 counts (42 sources). The normalisation, photon index
and column density were also free parameters. In seven cases, shown
in Figure~\ref{fig:abundance}, the abundances could be constrained.
These have more than 30000 photon counts each. We comment on some
individual sources in Figure~\ref{fig:abundance} in detail: We note
that \citet{Giuliani2014} found a metal-free solution ($Z/Z_{\odot}<0.05$)
when analysing a \emph{XMM-Newton} spectrum of GRB~120711A. The difference
may be because we use a slightly higher galactic column density of
$1.06\times10^{21}\textrm{cm}^{-2}$ following \citet{Evans2009}
and more recent absorption model and cross-sections in our fit. Our
abundance contours for GRB~090618 are consistent with those from
the \emph{XMM-Newton} spectral analysis of \citet{Campana2011}. They
could also place a lower limit of $Z/Z_{\odot}>0.2$. For GRB~130925A
and GRB~130907A we additionally excluded the (early-time) WT mode
spectra that may still be, despite our efforts of Section~\ref{sub:Data-reduction},
slightly contaminated by prompt emission. Nevertheless we obtain constraints
with the (late-time) PC mode data alone. For GRB~130925A, we note
that \citet{Schady2015} derived super-solar metallicity from optical
spectroscopy. Particular noteworthy is GRB~130702A. This LGRB is
associated with a supernova hosted by a dwarf galaxy with $M_{\star}\approx10^{8}M_{\odot}$
\citep{Kelly2013b}, and hence represents a typical LGRB. In all cases,
the constraints exclude a He-dominant absorber but are consistent
with the abundances of the local ISM.\nocite{Campana2011,Perley2011,Veres2015,Toy2016,Martin-Carrillo2014,Evans2014}

\begin{figure*}
\includegraphics[width=0.9\textwidth]{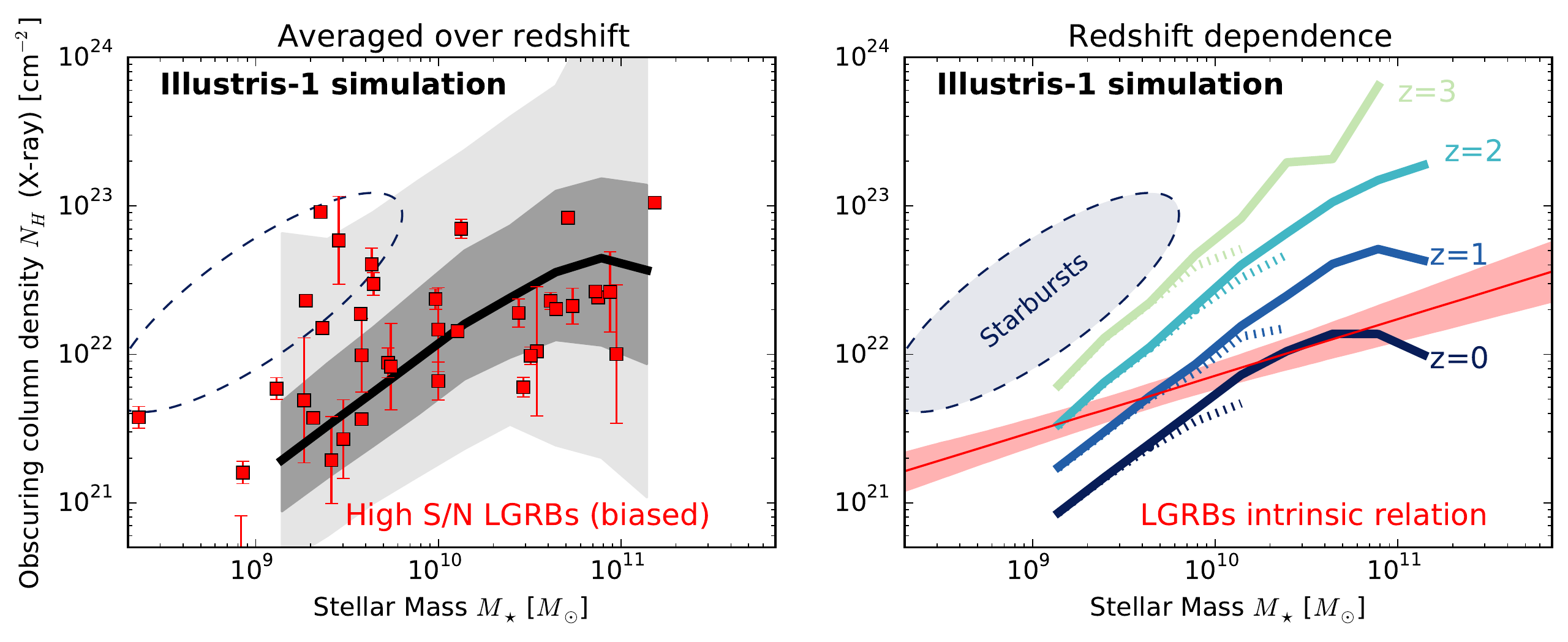}

\caption{\label{fig:sim-total}Obscuration in simulated galaxies. \emph{Left
panel}: The grey shaded regions show the 1 and 3 sigma range of the
distributions in median $\NH$ as seen from the densest region. The
black line marks the median. Individual observed LGRBs are overlaid
as red error bars (same as in Figure \ref{fig:MNH-rel}). While in
the left panel redshift snapshots have been combined according to
the SHOALS redshift distribution, the \emph{right panel} shows the
median as a function of redshift (thick lines). Dotted lines indicate
the effect of only using simulated galaxies with sub-solar stellar
metallicity. In red, our $\NH-M_{\star}$ relation is shown. The ellipse
indicates (for $z=0$) a sub-population in the simulation showing
elevated obscuring columns. These are predominantly star-burst galaxies.}
\end{figure*}
Additionally, if the LGRB is responsible for ionising substantial
fractions of the absorber, the effective absorbing column should be
reduced for more energetic bursts. \citet{Campana2012} find no significant
difference in the column distribution of bright bursts. In Figure~\ref{fig:energy-NH}
we show the distribution of sources in isotropic energy and effective
column density $\NH$ for sources with redshift information. The isotropic
energies were computed using the method of \citet{Bloom2003}\footnote{In a few cases where the peak energy could not be constrained, the
values were taken from \citet{Butler2010}.}. Effective column densities were derived using a local ISM abundance,
neutral absorber model. The \textsc{SPHERE} model is used, which is
valid also for the highest absorbing columns, as discussed in Section~\ref{sub:X-ray-spectral-analysis}.
There does not appear to be a burst energy-dependence in the absorber
properties. If the burst energy ionised metals and hydrogen, reducing
the effective column density by a factor of seven (example calculation
in W13, red arrow), we would expect a deficit of sources in the upper
right quadrant of the plot. However, no deficit of high-obscuration
sources at the luminous end is apparent in Figure~\ref{fig:energy-NH},
and the column density distribution appears independent of energy.
In Figure~\ref{fig:energy-NH}, only GRBs for which redshifts have
been determined are shown, which reduces the number of faint, obscured
bursts (lower right quadrant). To avoid such biases, black crosses
show LGRB from the SHOALS survey. Additionally, metals have been detected
in highly-absorbed, energetic bursts: Orange circles in Figure~\ref{fig:energy-NH}
show the same sources as in Figure~\ref{fig:abundance}. 

The correlation of column density and stellar mass dependence is a
strong argument that the X-ray column density is \emph{predominately}
due to the host galaxy-scale gas. The lack of energy-dependence of
the absorber support the dominance of a distant obscurer. Abundance
measurements suggest that the X-ray obscurer can be modelled similar
to the local ISM. Under local ISM abundances, the dominant absorbers
are Fe and O. Partial ionisation of metals may still be present and
account for deviations in $A_{V}$, but its effect on the X-ray spectrum
appears negligible. Furthermore, relatively low galaxy-scale column
densities can occur if the LOS does not pass through the galaxy (left
tail in Figure~\ref{fig:ellipsoid}, particularly objects below the
$M_{\star}-\NH$ relation in Figure~\ref{fig:MNH-rel}). In these
cases, the dominant obscurer could be the local environment, where
hard burst radiation destroying dust may reduce the $A_{V}/\NH$ ratio.
However, fully resolving the discrepancy between $\NH$ and $A_{V}$
measurements is beyond the scope of this work.

\subsection{Obscuration in simulated galaxies}

\label{sub:Obscuration-in-simulated}

At higher redshifts the gas content of galaxies is not easily accessibly
through observations. Instead we turn to simulated galaxies from hydrodynamic
cosmological simulations. This exercise is potentially predictive
because the amount of gas inside galaxies is constrained by the simulation's
requirement to start with the Big Bang's density and to reproduce
today's stellar mass function. In galaxy evolution models, the massive
end of the existing stellar population expels metals into the galaxy.
The metal gas produced per stellar mass is determined by the chosen
IMF and the metal yield, with the latter tuned to reproduce the stellar
mass function \citep{Lu2015}. The total metal gas mass residing in
galaxies further depends on the chosen feedback models, which can
expel gas out of the galaxy. Typically the metal gas mass inside galaxies
follows a $M_{Z}/M_{\star}=1:30-1:100$ relation in semi-analytic
models at $z=0-3$ \citep[e.g.][]{Croton2006,Croton2016}. The crucial
remaining question surrounds the arrangement of that gas inside galaxies,
as the concentration of gas defines its column density -- this requires
hydrodynamic simulations.

The \emph{Illustris} simulation \citep{Vogelsberger2014,Vogelsberger2014a}
is a cosmological hydrodynamic simulation which attempts to reproduce
the galaxy population using state-of-the-art star formation, supernovae
and AGN feedback mechanisms inside dark matter haloes. \emph{Illustris}
reproduces many observed quantities; most relevant for this work it
reproduces roughly the stellar mass distribution of galaxies, their
morphology, and gas content from CO observations \citep{Vogelsberger2014,Genel2014}.
The gas particle resolution in \emph{Illustris} is adaptive, with
some cells being as small as 48pc in the highest resolution simulation
(\emph{Illustris-1}) used here, indicating that today's cosmological
simulations indeed resolve galaxies into small substructures. 

We apply ray-tracing, treating each simulated galaxy (subhalo) separately.
The starting point is the densest region, presumably representing
a region of star formation. From that position, we radiate along random
sight-lines all metal gas bound to the subhalo. In \emph{Illustris},
gas is represented by Voronoi cells, therefore we Voronoi-tessellate
the ray and assign each part the corresponding cells density and finally
sum to a total metal column density. We then compute a equivalent
hydrogen column density distribution by converting under \citet{Wilms2000}
solar abundances to $\NH$. This mimics how $\NH$ is derived in X-ray
observations. We adopt $h=0.7$ and work in physical units at redshift
slices $z=0,\,1,\,2$ and $3$. We investigate all galaxy subhaloes
with $M_{\star}>10^{9}M_{\odot}$ (as smaller galaxies are difficult
to resolve).

The first question to address is whether the gas in simulated galaxies
reproduces the same $\NH$ values as observed. The left panel of Figure
\ref{fig:sim-total} shows several individual LGRBs from the SHOALS
sample, specifically those with host mass and $\NH$ measurements.
Overlaid are the results of the simulation snapshots, redshift-weighted
according to the SHOALS redshift distribution. Grey shading represents
the distribution of the median $\NH$ of individual simulated galaxies.
We find that the observations overlap well with the simulations under
the assumption that LGRBs originate in dense regions of galaxies.
We also find that the simulations predict a diversity of galaxies
with a scatter of $\sim0.5\,\text{dex}$ in $\NH$, in agreement with
our observations. We test the importance of the immediate vicinity
to the $\NH$ by excluding the inner $100\,\text{pc}$ radius, which
reduces $\NH$ by a factor of 2 on average. This indicates that distant,
i.e., galaxy-scale obscuration is important.

In both the observations and simulations, some rare objects occupy
the upper left quadrant of Figure \ref{fig:sim-total}, showing high
obscurations $\NH>10^{22}\text{cm}^{-2}$ despite low masses $M_{\star}<10^{10}M_{\odot}$.
In the simulations, these galaxies have high star formation rates,
with the majority being starburst galaxies and many having recently
experienced mergers. Figure \ref{fig:sim-total} indicates the relevant
range with an ellipse for $z=0$ (other redshifts are slightly higher).

The right panel of Figure \ref{fig:sim-total} depicts the redshift
evolution of simulated galaxies. In \emph{Illustris}, $z=1-3$ galaxies
have a slightly higher specific gas content, which affects the median
column densities. At low redshifts ($z=0-1$), massive galaxies in
particular loose gas by strong feedback from active galactic nuclei
to avoid over-production of massive galaxies. From our observations
we ruled out a redshift-dependent trend of LGRBs in Section \ref{sub:Host-mass-dependence}\footnote{The same result occurs when adding a redshift-dependence to the $\NH-M_{\star}$
relation; the individual constrained SHOALS LGRBs also do not appear
to follow that redshift evolution. }. Furthermore, at each redshift the slope of the $\NH-M_{\star}$
relation derived from \emph{Illustris} is substantially steeper than
$\frac{1}{3}$. 

These two issues indicate that the \emph{Illustris} simulation may
not represent the gas in galaxies correctly, because it predicts substantially
more obscured LGRBs and higher columns for LGRBs in massive galaxies.
If remaining concerns e.g., regarding absorber geometry and substructure
can be addressed, our X-ray tomography of galaxy gas could be used
in the future to distinguish (the strength of) feedback models and
star formation efficiencies (see also Paper II). Alternatively, the
lower observed obscuration may be due to a environmental preferences
of LGRBs inside galaxies, such as a metal/dust aversion.

\section{Summary}

\label{sec:Summary}We analysed a large sample of \emph{Swift}-detected
long-duration Gamma-Ray Bursts using modern statistical techniques,
incorporating the uncertainties from spectral analysis and investigating
the effect of redshift incompleteness from dust-extinct/dark LGRBs.
Our findings can be summarised as follows:
\begin{enumerate}
\item The column density of the LGRB population lies in the $10^{20-23}\text{cm}^{-2}$
range and can be described by a normal distribution.
\item A well-suited model for the column density distribution is a axisymmetric
ellipsoid of gas with randomly placed GRBs within. This set-up generalises
previous models based on the giant molecular clouds of the Milky Way.
Those in fact have lower column densities than observed from GRBs
in host galaxies of similar mass. Permitted solutions for the obscuring
clouds include a degenerate range of densities and flatness ($\sim$1:20).
Additionally it is necessary that the gas ellipsoid population has
a distribution in its total gas density of about $\Rellscat\,\text{dex}$.
\item We systematically search the \emph{Swift} archive for evidence of
heavily-obscured LGRBs. We note that such LGRB could have been detected
and characterised by \emph{Swift/}XRT given their intrinsic X-ray
luminosities, but are not observed. LGRBs therefore do not reach heavily
obscured column densities of $\NH>10^{23}\text{cm}^{-2}$.
\item The column density of LGRBs shows no significant evolution with redshift.
If present its effect is at most a factor of $\Revolfactor$.
\item LGRBs in galaxies of high stellar mass show substantially more obscuration.
We find a novel relation: 
\[
\NH=10^{21.7}\text{cm}^{-2}\times\left(M_{\star}/10^{9.5}M_{\odot}\right)^{1/3}
\]

\item The scatter in column densities can be fully explained by the mass-dependence
(v) and geometric effects (ii).
\item We argue based on the mass-dependence of the obscuration and the derived
geometry of the obscurer as well as analysis of well-mapped local
galaxies, that \textbf{the obscurer is predominantly the GRB host
galaxy itself}. 
\item This conclusion is corroborated by investigating the metal gas mass
in simulated galaxies. These predict the same magnitude of obscuring
X-ray column densities, similar scatter as well as a mass-dependence,
although of a steeper slope.
\end{enumerate}

\section*{Acknowledgements}

JB thanks David Rosario, Patricia Schady, Hendrik van Eerten, Jochen
Greiner, Antonis Georgakakis, Kirpal Nandra and Dave Alexander for
insightful conversations, and Pierre Maggi and Sergio Molinari for
insightful input on the obscuration in the Magellanic clouds and in
the CMZ. JB thanks Klaus Dolag, Sergio Contreras and Torsten Naab
for conversations about hydrodynamic simulations. 

We acknowledge support from the CONICYT-Chile grants Basal-CATA PFB-06/2007
(JB, FEB, SS), FONDECYT Regular 1141218 (SS, FEB), FONDECYT Postdoctorados
3160439 (JB) and 3140534 (SS), ``EMBIGGEN'' Anillo ACT1101 (FEB),
and the Ministry of Economy, Development, and Tourism's Millennium
Science Initiative through grant IC120009, awarded to The Millennium
Institute of Astrophysics, MAS (JB, FEB, SS).

We thank the builders and operators of \emph{Swift}. This work made
extensive use of data supplied by the UK Swift Science Data Centre
at the University of Leicester. This research has made use of software
provided by the Chandra X-ray Center (CXC) in the application package
CIAO and Sherpa. Additionally, the BXA\footnote{\url{https://johannesbuchner.github.io/BXA/}},
PyMultiNest\footnote{\url{https://johannesbuchner.github.io/ PyMultiNest/}},
Astropy \citep{AstropyCollaboration2013} and CosmoloPy\footnote{\url{http://roban.github.com/CosmoloPy/}}
software packages were used. 

\bibliographystyle{mnras}
\bibliography{agn,grb,sim,gcn}

\appendix

\section{Photon Index Distribution}

\begin{table}
\caption{\label{tab:Photon-Index-fits}Photon index Gaussian fits.}

\centering{}\input{gammatable.tex}
\end{table}

\label{sec:Gamma-Luminosity}For completeness, we report the intrinsic
photon index distribution of GRB afterglows, which is a side-products
of our analysis. The X-ray spectrum of the complete sample was analysed
to obtain constraints on the slope of the intrinsic powerlaw. The
constraints for the complete sample is shown in Figure~\ref{fig:Caterpillar-plots}.
We approximate the distribution by a Gaussian distribution, plotted
in red. The overall distribution is centred at $\Gamma=1.94$ with
a width of $\sigma=0.20$. Table~\ref{tab:Photon-Index-fits} shows
the mean and standard deviations obtained from the various samples.
The Gaussian distribution provides a good approximation in the bulk
of the population. However the data show heavier tails, including
$\Gamma<1.5$ and $\Gamma>2.4$. The individual spectra show no obvious
signs of bad fits in either extreme. The high-mass subsample and redshift
subsample show steeper slopes with a narrower distribution, but the
difference is smaller than the uncertainties. 
\begin{figure}
\begin{centering}
\includegraphics[width=0.99\columnwidth]{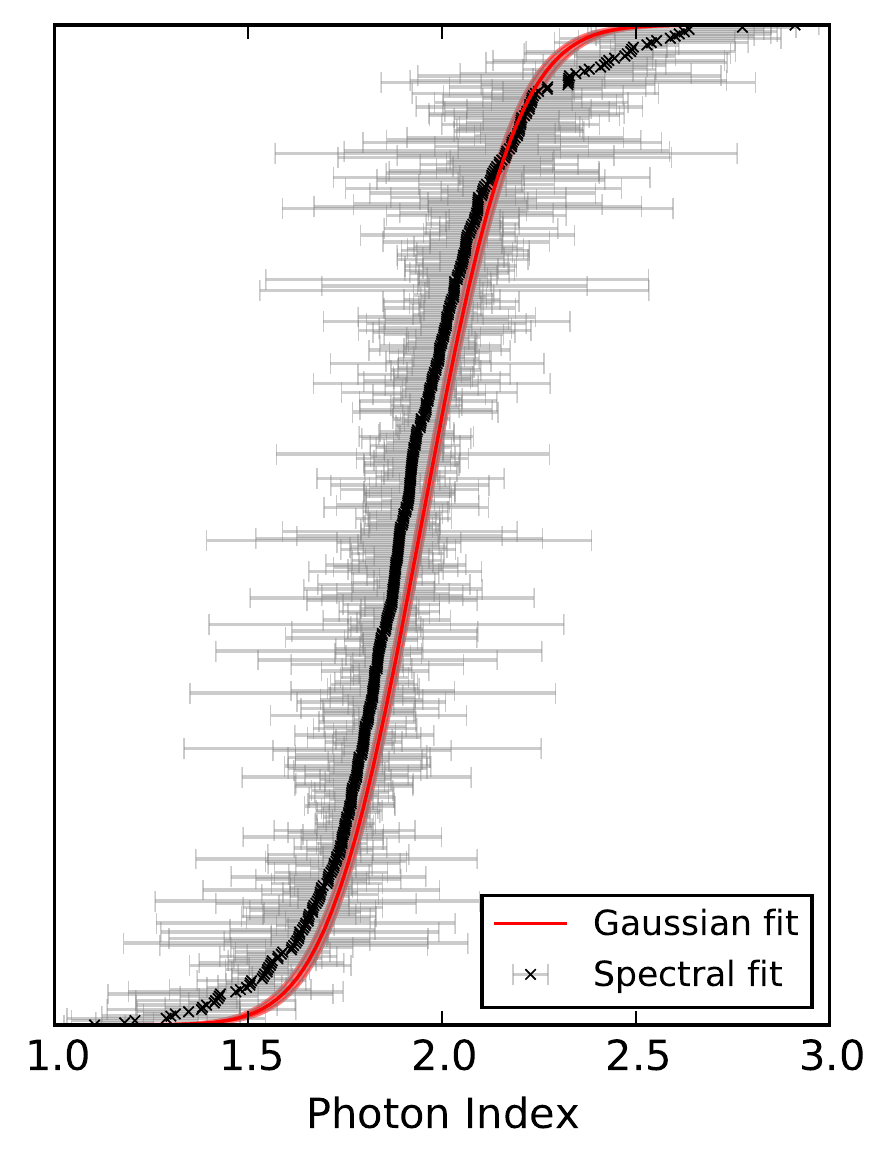}
\par\end{centering}

\caption[Caterpillar plot of Photon Index]{\label{fig:Caterpillar-plots}Caterpillar plot of the photon index.
Constraints on the photon index of the intrinsic powerlaw are shown
as error bars for each object, sorted by their mean. During the population
analysis we assumed a Gaussian for the distribution of photon indices
(red). The spectral index $\beta$ is related to the photon index
as $\beta=\Gamma-1$.}
\end{figure}

\citet{Curran2010} investigate the photon index distribution based
on the automatic fitting results of \citet{Evans2009} of 301 GRBs.
They find a peak at $\Gamma\approx2.1$, and the distribution spreads
the full range of $1-3$. In contrast, we find that the standard deviation
of the distribution is much narrower ($\sigma=0.2$) than their analysis
suggests ($\sigma\approx0.5$). This is probably because of their
use of best-fit values, which introduce additional scatter, and the
fact that the completely automated analysis of \citet{Evans2009}
sometimes chooses time windows affected by prompt emission while we
manually verified each time window. \citet{Wang2015} performed temporally
resolved fitting of X-ray and optical data and found $\Gamma=1.98$
with standard deviation $\sigma=0.15$, which is consistent with our
results.

\citet{Curran2010} then considers two regimes in which synchrotron
radiation is produced: (a) The electron energy distribution index
$p$ relates to the spectral index ($\beta=\Gamma-1$) as $p=2\cdot\beta$
when the cooling frequency is below the X-ray frequency ($\nu_{c}<\nu_{X}$),
and (b) as $p=2\cdot\beta+1$ otherwise ($\beta=(p[-1])/2+1$. Adopting
a Gaussian distribution for $p$ centred at, e.g., $2.2$ then yields
a double-peaked distribution for the photon index centred at $\Gamma=2.1$,
with a secondary, small component (case b) offset by $0.5$. Here,
we find additional contributions at both lower and higher photon indices
which necessitate a different model than two Gaussian peaks, and we
can therefore not unambiguously conclude that $\nu_{c}>\nu_{X}$ applies
in most cases. Nevertheless, if that is assumed, we find $p=2$ for
the electron density distribution index distribution with a standard
deviation of $0.4$.

\section{X-ray background model}

\label{sec:bkg-model}The shape of the XRT background has been analysed
by \citet{Pagani2007} (see Figure 2 there). It shows a steep increase
below $0.5\,\text{keV}$ and several bumps. The background spectra
analysed in this work show the same shape. We fit the background spectrum
with a broken powerlaw model and four Gaussian components at $\sim0.7,\,2.2,\,1.2$
and $0.4\text{\,keV}$, by order of importance. The parameters of
this model are optimised according to the Poisson likelihood. In further
analysis, the background model parameters are held fixed, and the
background model is added to the source spectral fit, scaled by the
area ratio of the spectral extraction regions. Our code and model
for fitting the \emph{Swift/}XRT background is available as part of
the BXA software.

\section{Numerical Details on Fitting the Ellipsoids models}

\label{sec:Numerical-Details}This section describes how the SingleEllipsoid
model is computed. Monte Carlo ray-tracing simulations are used to
compute the column density distribution $p(\NH|M,\theta)$.

First, random points inside the ellipsoid are generated. The ellipsoids
equation,

\begin{equation}
\left(\frac{p_{x}}{r_{x}}\right)^{2}+\left(\frac{p_{y}}{r_{y}}\right)^{2}+\left(\frac{p_{z}}{r_{z}}\right)^{2}\leq1,\label{eq:ellipsoid}
\end{equation}
describes whether a point $\mathbf{p}=(p_{x},p_{y},p_{z})$ is inside.
The radii are $r_{x}=r_{y}=R$ and $r_{z}=z$ under cylindrical symmetry.
For constant density sampling, $(p_{x},p_{y},p_{z})$ are first drawn
uniformly in $p_{x}\sim U(-R,R)$, $p_{y}\sim U(-R,R)$, $p_{z}\sim U(-z,z)$
and the vector $\mathbf{p}$ is rejected if outside the ellipsoid. 

Second, a random unit direction vector is generated. Three unit normal
variates are combined to a vector $\mathbf{d}\sim(N(0,1),\,N(0,1),\,N(0,1))$
which is then normalised to unit length $\mathbf{n}=\mathbf{d}/|\mathbf{d}|$. 

Third, the length $l$ of the ray inside the ellipsoid needs to be
computed, which is the distance between $\mathbf{p}$ and the point
where the ray exits the ellipsoid, $\mathbf{q}$. The coordinates
of $\mathbf{q}$ are governed by Equation \ref{eq:ellipsoid} and
the line equation,

\begin{equation}
\mathbf{q}=\mathbf{p}+l\cdot\mathbf{n},\label{eq:line}
\end{equation}
which yield the quadratic equation
\begin{equation}
\left(\frac{p_{x}+l\cdot n_{x}}{r_{x}}\right)^{2}+\left(\frac{p_{y}+l\cdot n_{y}}{r_{y}}\right)^{2}+\left(\frac{p_{z}+l\cdot n_{z}}{r_{z}}\right)^{2}=1\label{eq:quad}
\end{equation}
with $l$ unknown. Equation \ref{eq:quad} has two solutions for $l$
(one for the positive, one for the negative direction). Only the positive
one is considered. With 

\begin{eqnarray}
b & = & \frac{p_{x}\cdot n_{x}}{r_{x}^{2}}+\frac{p_{y}\cdot n_{y}}{r_{y}^{2}}+\frac{p_{z}\cdot n_{z}}{r_{z}^{2}}\\
a & = & \frac{n_{x}^{2}}{r_{x}^{2}}+\frac{n_{y}^{2}}{r_{y}^{2}}+\frac{n_{z}^{2}}{r_{z}^{2}}\\
c & = & \frac{p_{x}^{2}}{r_{x}^{2}}+\frac{p_{y}^{2}}{r_{y}^{2}}+\frac{p_{z}^{2}}{r_{z}^{2}}\\
d & = & b^{2}-a\cdot c\\
l & = & \begin{cases}
0 & \text{if}\,d<0\\
(-b+\sqrt{d})/a & \text{otherwise}
\end{cases},
\end{eqnarray}
we can finally write the column probed by the ray as $\NH=\NH^{\text{major}}\cdot l$,
where $\NH^{\text{major}}$ is the column density of the ellipsoid
in a unit length. The problem is fundamentally degenerate ($\NH^{\text{major}}$
and size), so $R=1$ is assumed for the SingleEllipse model.

To compute the column density distribution, $p(\NH|M,\theta)$, $400000$
random rays are generated. A histogram of their $\NH$ between $10^{19}-10^{26}\text{cm}^{-2}$
with 100 logarithmically spaced bins provides a well-sampled approximation.
A problem may occur with less obscured column densities in the simulation,
which can never be measured due to Milky Way absorption. For simplicity,
the column density of each ray is modified as $\NH'=\NH+10^{u}$ with
the random number $u\sim U(19,20)$ to ensure all rays have $\NH>10^{19}\text{cm}^{-2}$.
This redistributes unobscured rays to the range $10^{19}-10^{20}\text{cm}^{-2}$,
uniformly, and, although done primarily for numerical reasons, may
be interpreted as placing the ellipsoid in a low-density gas with
$\NH<10^{20}\text{cm}^{-2}$. 

The arising distribution (see Figure~\ref{fig:ellipsoid}) cannot
be approximated by simple analytic formulas. Towards low $\NH$ values,
the distribution rises exponentially. For very low $z/R$ ratios,
the distribution declines exponentially toward high $\NH$ values,
but the peak is too wide/narrow to be fitted by a broken/bending powerlaw.
Additionally, for moderate $z/R$ ratios, there is a steep truncation
at $\NH=2$ (see Figure~\ref{fig:ellipsoid}) which declines faster
than a exponential cut-off. 

To emulate a dispersion in the population of $\log\NH^{\text{major}}$
of standard deviation $\sigma$, the histogram is convolved with a
Gaussian. Finally, linear interpolation of the histogram is used to
evaluate $p(\NH|M,\theta)$ at arbitrary $\NH$ values.

The MultiEllipsoid model generates points proportional to the mass
in each ellipsoid, which is 
\[
M\propto\NH^{\text{major}}\cdot r_{x}\cdot r_{y}\cdot r_{z}
\]
The ellipsoid to draw from is chosen randomly in proportion to its
masses $M$. Rays now may probe multiple ellipsoids, and the final
$\NH$ is the sum of all ellipsoids encountered. The definition of
$l$ has to be modified because some rays may not originate inside
the ellipsoid at hand (checked with Equation \ref{eq:ellipsoid}),
but cross it. In that case the distance $l$ is between the two quadratic
solutions, giving

\[
l_{\text{cross}}=2\cdot\sqrt{d}/a.
\]
Otherwise, the same procedure as in the SingleEllipsoid model is applied.

To simulate the column density from the centre, $\mathbf{p}=(0,0,0)$
is set fixed, and the procedure of generating random rays is applied
in the same fashion.

\bsp	
\label{lastpage}
\end{document}

%% file: stats.tex
\newcommand{\nhlimit}{21}
\newcommand{\blimit}{20}
\newcommand{\nlong}{844}
\newcommand{\ntotal}{920}

\newcommand{\nlongbrangewithz}{208}
\newcommand{\nlongbrange}{512}

\newcommand{\nlongbrangeinzrangesecureobsc}{28}
\newcommand{\nlongbrangeinzrange}{163}

%% file: resultvalues.tex
\newcommand{\Rellscat}{0.22}
\newcommand{\Rellscaterr}{0.14}
\newcommand{\Rellscatqlo}{0.03}

\newcommand{\Rgaussmean}{21.6}
\newcommand{\Rgaussscat}{0.6}

\newcommand{\Revolfactor}{3}

\newcommand{\RMlinenorm}{21.7}

\newcommand{\RMlineslope}{0.38}

\newcommand{\RMlinenormfine}{21.67}
\newcommand{\RMlinenormerrfine}{0.06}
\newcommand{\RMlinescatfine}{0.49}
\newcommand{\RMlinescaterrfine}{0.05}
\newcommand{\RMlineslopefine}{0.38}
\newcommand{\RMlineslopeerrfine}{0.06}

%% file: gcncite.tex
\nocite{2016GCN..18982...1P,2016GCN..18965...1M,2016GCN..18966...1D,2016GCN..18969...1X,2016GCN..18925...1S,2016GCN..18886...1D,2016GCN..18915...1D,2015GCN..18696...1X,2015GCN..18603...1B,2015GCN..18598...1B,2015GCN..18540...1M,2015GCN..18524...1T,2015GCN..18505...1X,2015GCN..18506...1X,2015GCN..18483...1C,2015GCN..18487...1P,2015GCN..18493...1Z,2015GCN..18426...1D,2015GCN..18318...1D,2015GCN..18273...1Z,2015GCN..18274...1D,2015GCN..18187...1D,2015GCN..18177...1S,2015GCN..17758...1C,2015GCN..17755...1M,2015GCN..17697...1S,2015GCN..17710...1D,2015GCN..17672...1P,2015GCN..17616...1P,2015GCN..17583...1D,2015GCN..17523...1D,2015GCN..17420...1K,2015GCN..17358...1C,2015GCN..17278...1C,2014GCN..17234...1G,2014GCN..17228...1P,2014GCN..17198...1D,2014GCN..17177...1C,2014GCN..17177...1C,2014GCN..17081...1P,2014GCN..17040...1X,2014GCN..16968...1D,2014GCN..16891...1S,2014GCN..16902...1D,2014GCN..16797...1C,2014GCN..16774...1C,2014GCN..16570...1T,2014GCN..16505...1C,2014GCN..16437...1H,2014GCN..16437...1H,2014GCN..16401...1K,2014GCN..16301...1C,2014GCN..16269...1C,2014GCN..16310...1D,2014GCN..16217...1F,2014GCN..16194...1K,2014GCN..16181...1P,2015GCN..17988...1J,2014GCN..15961...1T,2014GCN..15964...1D,2014GCN..15966...1C,2014GCN..15922...1J,2014GCN..15924...1D,2014GCN..15936...1J,2014GCN..15831...1S,2013GCN..15624...1C,2013GCN..15493...1T,2013GCN..15494...1H,2013GCN..15451...1X,2013GCN..15445...1C,2013GCN..15407...1X,2013GCN..15408...1D,2013GCN..15307...1C,2013GCN..15310...1D,2013GCN..15187...1D,2013GCN..15144...1C,2013GCN..14956...1X,2013GCN..14882...1T,2013GCN..14848...1S,2013GCN..14744...1T,2013GCN..14747...1S,2013GCN..14748...1C,2013GCN..14685...1S,2013GCN..14687...1C,2013GCN..14621...1C,2013GCN..14493...1F,2013GCN..14455...1L,2013GCN..14478...1X,2013GCN..14491...1F,2013GCN..14605...1G,2013GCN..14617...1W,2013GCN..14669...1S,2013GCN..14617...1W,2013GCN..14437...1D,2013GCN..14380...1D,2013GCN..14390...1K,2012GCN..13929...1T,2012GCN..13930...1K,2012GCN..13890...1T,2012GCN..13903...1L,2012GCN..13810...1K,2012GCN..13730...1H,2012GCN..13723...1S,2012GCN..13477...1F,2012GCN..13457...1E,2012GCN..13458...1T,2012GCN..13460...1X,2012GCN..13348...1T,2012GCN..13251...1T,2012GCN..13257...1S,2012GCN..13133...1P,2012GCN..13134...1K,2012GCN..13146...1S,2012GCN..12865...1C,2012GCN..12866...1P,2012GCN..12867...1M,2011GCN..12648...1V,2013GCN..14273...1X,2013GCN..14273...1X,2011GCN..12429...1L,2011GCN..12431...1W,2011GCN..12284...1D,2011GCN..11993...1D,2011GCN..11997...1D,2011GCN..11635...1D,2011GCN..11638...1C,2011GCN..11640...1V,2011GCN..11579...1D,2011GCN..11575...1K,2011GCN..11518...1C,2010GCN..11464...1P,2010GCN..11468...1C,2010GCN..11469...1C,2011GCN..11518...1C,2010GCN..11230...1T,2010GCN..11195...1C,2010GCN..11026...1O,2010GCN..11317...1F,2010GCN..10971...1T,2010GCN..10752...1C,2010GCN..10620...1A,2010GCN..10624...1C,2010GCN..10621...1M,2010GCN..10495...1V,2010GCN..10439...1K,2010GCN..10441...1G,2010GCN..10443...1C,2010GCN..10445...1D,2010GCN..10433...1B,2010GCN..10432...1H,2009GCN..10202...1C,2009GCN..10233...1T,2009GCN..10065...1C,2009GCN..10093...1C,2009GCN..10038...1C,2009GCN..10042...1D,2009GCN..9771....1D,2009GCN..9761....1M,2009GCN..9264....1L,2009GCN..9269....1T,2008GCN..8711....1P,2008GCN..8601....1L,2008GCN..8438....1D,2008GCN..8448....1C,2008GCN..8335....1B,2008GCN..8301....1V,2008GCN..8304....1C,2008GCN..8212....1H,2008GCN..8191....1V,2008GCN..7962....1P,2008GCN..7601....1V,2008GCN..7615....1C,2008GCN..7388....1P,2008GCN..7389....1B,2008GCN..7391....1V,2008GCN..7397....1P,2007GCN..7152....1D,2007GCN..7154....1B,2007GCN..7151....1B,2007GCN..6741....1T,2007GCN..6665....1C,2006GCN..5617....1J,2006GCN..5573....1L,2006GCN..5555....1R,2006GCN..5674....1T,2006GCN..5535....1V,2006GCN..5513....1F,2007GCN..6663....1T,2006GCN..5456....1M,2006GCN..5457....1C,2007GCN..6663....1T,2006GCN..5373....1T,2006GCN..5320....1J,2006GCN..5337....1J,2006GCN..5319....1J,2006GCN..5298....1J,2006GCN..5275....1P,2006GCN..5276....1F,2006GCN..5277....1F,2006GCN..5282....1C,2006GCN..5237....1L,2006GCN..5223....1P,2006GCN..5226....1S,2006GCN..5489....1F,2007GCN..6166....1S,2006GCN..5489....1F,2006GCN..5218....1C,2007GCN..6997....1J,2006GCN..5170....1B,2006GCN..5155....1C,2006GCN..5217....1B,2006GCN..5156....1C,2006GCN..5217....1B,2006GCN..5104....1P,2006GCN..5123....1O,2006GCN..5123....1O,2006GCN..5161....1T,2006GCN..5238....1B,2006GCN..5071....1B,2006GCN..5077....1P,2006GCN..5072....1H,2006GCN..5238....1B,2006GCN..5052....1C,2006GCN..4969....1D,2006GCN..5002....1P,2006GCN..4974....1V,2006GCN..4815....1B,2006GCN..4792....1M,2006GCN..4803....1M,2006GCN..4804....1S,2006GCN..4809....1F,2006GCN..5376....1F,2006GCN..4783....1M,2006GCN..4729....1C,2006GCN..4753....1H,2006GCN..4686....1F,2006GCN..4692....1F,2006GCN..4701....1P,2006GCN..4703....1A,2006GCN..4591....1M,2006GCN..4592....1C,2006GCN..4593....1P,2006GCN..4622....1B,2006GCN..4545....1G,2006GCN..4583....1P,2006GCN..4520....1P,2006GCN..4539....1M,2005GCN..4384....1B,2005GCN..4375....1S,2005GCN..4291....1T,2005GCN..4255....1H,2005GCN..4271....1P,2006GCN..5387....1P,2006GCN..5387....1P,2005GCN..4221....1Q,2005GCN..4186....1S,2005GCN..4017....1J,2005GCN..4029....1J,2005GCN..4032....1P,2005GCN..4044....1D,2005GCN..3948....1F,2005GCN..3949....1F,2005GCN..3971....1P,2006GCN..5982....1H,2006GCN..4749....1H,2005GCN..3874....1F,2005GCN..3833....1P,2005GCN..3860....1L,2005GCN..3809....1J,2005GCN..3758....1B,2005GCN..3753....1B,2005GCN..3749....1F,2005GCN..3709....1C,2005GCN..3732....1P,2005GCN..3710....1R,2005GCN..3746....1D}

%% file: cat-excerpt.tex
\begin{tabular}{l c c c c c c c c c c}
Name & Duration & RA & Dec & b{[}\textdegree {]} & $N^\text{MW}_\text{H}$ & z & Time selection[s] & $N_\text{H}$ & $N_{\text{H},10\%}$ & $N_{\text{H},90\%}$ \tabularnewline
\hline
GRB 041218 & -1 & 01:39:07 & +71:20:29 & 8.8 & ${4.17}\times{10}^{21}$ & 0 & 19350-20616 & ${8.55}\times{10}^{20}$ & ${1.85}\times{10}^{19}$ & ${5.30}\times{10}^{22}$ \tabularnewline 
GRB 061021 & 46.2 & 09:40:36 & -21:57:05 & 22.6 & ${5.53}\times{10}^{20}$ & 0.3463 & 367-4384111 & ${2.01}\times{10}^{20}$ & ${1.09}\times{10}^{20}$ & ${4.92}\times{10}^{20}$ \tabularnewline 
GRB 061021 & 46.2 & 09:40:36 & -21:57:05 & 22.6 & ${5.53}\times{10}^{20}$ & 0.3463 & 367-4384111 & ${2.01}\times{10}^{20}$ & ${1.09}\times{10}^{20}$ & ${4.92}\times{10}^{20}$ \tabularnewline 
GRB 080207 & 340 & 13:50:03 & +07:30:08 & 66.0 & ${2.11}\times{10}^{20}$ & 2858 & 130-324571 & ${1.06}\times{10}^{23}$ & ${8.97}\times{10}^{22}$ & ${1.24}\times{10}^{23}$ \tabularnewline 
GRB 150616A & 599.5 & 20:58:52 & -53:23:38 & -40.3 & ${3.38}\times{10}^{20}$ & 0 & 7635-312789 & ${2.69}\times{10}^{22}$ & ${9.98}\times{10}^{21}$ & ${1.00}\times{10}^{23}$ \tabularnewline 
 ...  &  &  &  &  &  &  &  &  &  \tabularnewline
\end{tabular}

%% file: paramtable.tex
\begin{tabular}{ccccccc}
\textbf{Model} & Component & \multicolumn{3}{c}{Parameters} & $\mathbf{N}_{\text{params}}$ & \textbf{$\Delta$AIC}\tabularnewline
\hline 
\multicolumn{2}{l}{Broken powerlaw} & Break $N_{H}$ & Low $N_{H}$ slope & High $N_{H}$ slope & 5 & 0.0\tabularnewline
 &  & $\log a=21.87\pm0.06$ & $b=0.78\pm0.09$ & $c=-1.39\pm0.16$ &  & \tabularnewline
 &  &  &  &  &  & \tabularnewline
\multicolumn{2}{l}{Gaussians} & Mean & Standard deviation &  & 4 & -2.3\tabularnewline
 &  & $\mu=21.64\pm0.03$ & $\sigma=0.55\pm0.03$ &  &  & \tabularnewline
 &  &  &  &  &  & \tabularnewline
\multicolumn{2}{l}{Broken Powerlaws, 5 redshift bins} &  &  &  & 17 & 7.2\tabularnewline
 &  & Break $N_{H}$ & Low $N_{H}$ slope & High $N_{H}$ slope &  & \tabularnewline
 & $z<0.3$   & $\log a_{1}=20.67\pm0.25$ & $b_{1}=2.44\pm1.13$ & $c_{1}=-2.35\pm1.14$ &  & \tabularnewline
 & $z=0.3-1$ & $\log a_{2}=21.81\pm0.18$ & $b_{2}=0.74\pm0.18$ & $c_{2}=-1.85\pm0.70$ &  & \tabularnewline
 & $z=1-2$   & $\log a_{3}=21.83\pm0.09$ & $b_{3}=1.02\pm0.22$ & $c_{3}=-1.24\pm0.18$ &  & \tabularnewline
 & $z=2-4$   & $\log a_{4}=22.03\pm0.12$ & $b_{4}=0.68\pm0.12$ & $c_{4}=-1.67\pm0.44$ &  & \tabularnewline
 & $z>4$     & $\log a_{5}=21.05\pm0.61$ & $b_{5}=2.19\pm1.15$ & $c_{5}=-1.48\pm1.16$ &  & \tabularnewline
 &  &  &  &  &  & \tabularnewline
\multicolumn{2}{c}{Gaussians, 5 redshift bins} &  &  &  & 12 & -13.9\tabularnewline
 &  & Mean & Standard deviation &  &  & \tabularnewline
 & $z<0.3$   & $\mu_{1}=20.55\pm0.16$ & $\sigma_{1}=0.14\pm0.26$ &  &  & \tabularnewline
 & $z=0.3-1$ & $\mu_{2}=21.42\pm0.09$ & $\sigma_{2}=0.54\pm0.08$ &  &  & \tabularnewline
 & $z=1-2$   & $\mu_{3}=21.77\pm0.05$ & $\sigma_{3}=0.50\pm0.04$ &  &  & \tabularnewline
 & $z=2-4$   & $\mu_{4}=21.63\pm0.10$ & $\sigma_{4}=0.61\pm0.08$ &  &  & \tabularnewline
 & $z>4$     & $\mu_{5}=21.20\pm0.40$ & $\sigma_{5}=0.55\pm0.28$ &  &  & \tabularnewline
\end{tabular}

%% file: samplestable.tex
\begin{tabular}{ccc}
Parameter & Mean $\log\NH$ & Std. deviation $\sigma$\tabularnewline
\hline 
Complete sample & $21.64\pm0.03$ & $0.55\pm0.03$\tabularnewline
Low-mass sub-sample & $21.44\pm0.09$ & $0.65\pm0.07$\tabularnewline
High-mass sub-sample & $22.13\pm0.10$ & $0.43\pm0.09$\tabularnewline
Redshift sub-sample & $21.49\pm0.06$ & $0.58\pm0.05$\tabularnewline
\end{tabular}

%% file: gcncite2.tex
\nocite{2005GCN..3201....1B,2005GCN..3518....1G,2005GCN..4150....1G,2006GCN..5907....1O,2007GCN..6230....1G,2007GCN..6459....1G,2007GCN..6403....1G,2007GCN..6024....1H,2007GCN..7081....1K,2008GCN..7784....1G,2008GCN..8694....1G,2009GCN..9139....1C,2009GCN..9647....1G,2009GCN..9196....1P,2012GCN..13437...1G,2013GCN..14972...1C,2014GCN..15670...1G,2014GCN..16495...1G,2014GCN..16798...1Z,2015GCN..17579...1Y}
\nocite{2005GCN..3179....1G,2005GCN..4030....1G,2005GCN..4238....1G,2006GCN..4989....1G,2006GCN..5264....1G,2006GCN..5460....1G,2006GCN..5748....1G,2006GCN..5837....1G,2006GCN..5984....1P,2007GCN..6849....1G,2007GCN..6960....1G,2008GCN..7487....1G,2008GCN..7589....1G,2008GCN..7854....1G,2008GCN..8101....1S,2008GCN..8259....1G,2008GCN..8548....1G,2008GCN..8548....1G,2008GCN..8715....1B,2009GCN..8776....1G,2009GCN..9230....1C,2009GCN..9415....1M,2009GCN..9821....1P,2009GCN..10040...1M,2009GCN..10057...1G,2009GCN..10266...1M,2011GCN..11971...1G,2011GCN..12008...1G,2012GCN..13469...1G,2012GCN..13642...1J,2012GCN..13721...1Y,2013GCN..14487...1G,2013GCN..14808...1G,2013GCN..14958...1G,2013GCN..15145...1G,2013GCN..15203...1G,2013GCN..15413...1G,2013GCN..15455...1F,2014GCN..15790...1V,2014GCN..15833...1Z,2014GCN..16152...1V,2014GCN..16220...1J,2014GCN..16262...1S,2015GCN..17674...1Z,2009GCN..9553....1G,2009GCN..9679....1G,2009GCN..9756....1B,2010GCN..10670...1K,2010GCN..10882...1G,2010GCN..11119...1G,2010GCN..11169...1S,2010GCN..11248...1G,2011GCN..11659...1G,2011GCN..11723...1G,2011GCN..12166...1G,2011GCN..12223...1G,2011GCN..12276...1S,2011GCN..12433...1G,2012GCN..12872...1G,2012GCN..13137...1K,2012GCN..13910...1B,2012GCN..14010...1G,2013GCN..14368...1G,2013GCN..14417...1G}

%% file: gammatable.tex
\begin{tabular}{ccc}
Parameter & Mean & Standard deviation\tabularnewline
\hline 
Complete sample & $1.95\pm0.01$ & $0.19\pm0.01$\tabularnewline
Redshift sub-sample & $1.93\pm0.02$ & $0.19\pm0.01$\tabularnewline
SHOALS & $1.95\pm0.02$ & $0.23\pm0.02$\tabularnewline
low-mass sub-sample & $1.95\pm0.03$ & $0.25\pm0.02$\tabularnewline
high-mass sub-sample & $1.98\pm0.04$ & $0.18\pm0.04$\tabularnewline
\end{tabular}